\documentclass[12pt]{article}
\usepackage[dvips]{graphicx}
\usepackage{amssymb}

\makeatletter
 \@addtoreset{equation}{section}
 
\makeatother

\newcommand{\palpha}{\alpha'}

\newcommand{\maru}{\mbox{\tiny$\stackrel{\circ}{\scriptstyle\circ}$}}

\newcommand{\ww}{\mathtt{w}}
\newcommand{\nn}{\mathtt{n}}

\newcommand{\bphi}{\vec{\phi}}
\newcommand{\bchi}{\vec{\chi}}

\begin{document}


\begin{titlepage}

\renewcommand{\thefootnote}{\fnsymbol{footnote}}

\begin{flushright}
\begin{tabular}{l}
UTHEP-518\\
hep-th/0603152 \\
March 2006
\end{tabular}
\end{flushright}

\bigskip

\begin{center}
{\Large \bf D-branes and Closed String Field Theory}
\end{center}

\bigskip

\begin{center}
{\large
Yutaka Baba}${}^{a}$\footnote{e-mail:
        yutaka@het.ph.tsukuba.ac.jp},
{\large Nobuyuki Ishibashi}${}^{a}$\footnote{e-mail:
        ishibash@het.ph.tsukuba.ac.jp},
{\large Koichi Murakami}${}^{b}$\footnote{e-mail:
        murakami@het.ph.tsukuba.ac.jp}
\end{center}

\begin{center}
${}^{a}${\it
Institute of Physics, University of Tsukuba,\\
Tsukuba, Ibaraki 305-8571, Japan}\\
\end{center}
\begin{center}
${}^{b}${\it Center for Computational Sciences, University of Tsukuba,\\
Tsukuba, Ibaraki 305-8577, Japan}
\end{center}

\bigskip

\bigskip

\bigskip

\begin{abstract}
We construct BRST invariant solitonic states in the $OSp$ invariant 
string field theory for closed bosonic strings. Our construction
is a generalization of the one given in the noncritical case. 
These states are made by using the boundary states for D-branes, and 
can be regarded as states in which D-branes or ghost D-branes are
excited. We calculate the vacuum amplitude in the presence of solitons
perturbatively and show that the cylinder amplitude for the D-brane
is reproduced. The results imply that these are states with even number
of D-branes or ghost D-branes. 
\end{abstract}

\setcounter{footnote}{0}
\renewcommand{\thefootnote}{\arabic{footnote}}

\end{titlepage}

\section{Introduction}
D-branes have been studied for many years and used to reveal 
nonperturbative aspects of string theory. They are considered to be 
solitons in string theory.
{}From the viewpoint of open string theory, (for example,
in the vacuum string field theory \cite{Rastelli:2000hv}),
D-branes emerge
as soliton-like solutions of the equation of motion.

The question we would like to address in this paper is ``what are 
D-branes in closed string field theory?". Although several attempts 
have been made \cite{Hashimoto:1997vz}\cite{Asakawa:2003jv}, 
D-branes have not been studied so much in the context of closed 
string field theories. 

Actually, a fairly clear answer to the above question is given 
for noncritical strings. D-branes in noncritical string theories 
can be defined as in the critical ones \cite{Fateev:2000ik}. In 
Ref.\cite{Fukuma:1996hj}, Fukuma and Yahikozawa 
showed that the D-branes can be 
realized as solitonic operators which commute with the Virasoro 
and $W$ constraints \cite{Fukuma:1990jw} for the noncritical 
string theories.
In Ref.\cite{Hanada:2004im}, it was shown that how 
such solitonic operators are realized in the string field theory 
of noncritical strings presented in Ref.\cite{Ishibashi:1993pc}. 
States in which D-branes are excited can be made by acting 
these solitonic operators on the vacuum. 

What we would like to do in this paper is to construct such 
states in a critical bosonic string field theory. 
Since the string field 
Hamiltonian given in Ref.\cite{Ishibashi:1993pc} consists of the 
joining-splitting interactions, it seems possible to generalize 
the construction of the operators given in Ref.\cite{Hanada:2004im} to 
the string field theories with the light-cone gauge type interactions. 
In this paper, we 
take the $OSp$ invariant string field
theory \cite{Siegel:1984ap} as such a theory. 
The $OSp$ invariant string field theory is a covariantized version 
of the light-cone gauge string field theory
\cite{Kaku:1974zz}
and it was proved that 
the S-matrix elements coincide with those of the light-cone gauge 
one \cite{Kugo:1987rq}. 

What we will do is to construct solitonic operators made from 
the creation and annihilation operators of second-quantized strings. 
In order to deal 
with D-branes, we consider the closed strings
whose wave functions are proportional to  
the boundary states. Such states were shown \cite{Kishimoto:2003ru} 
to satisfy the idempotency equations. Because of these relations, we 
expect that three string vertices for such strings look quite like 
the ones which appear in the noncritical string field theory in 
Ref.\cite{Ishibashi:1993pc}. 
We will show that we can construct solitonic operators using such 
states. These operators can be considered as creation  
operators of D-branes
\footnote{In a recent paper\cite{Yoneya:2005si}, 
the author speculated about such operators 
from a quite different point of view. }
(or ghost D-branes recently proposed in Ref.\cite{Takayanagi}).
Acting them on the vacuum, we obtain BRST invariant states, which 
can be regarded as states in which D-branes are excited. 
We calculate the vacuum 
amplitude and show that these operators create two D-branes. 

The organization of this paper is as follows.
In section \ref{sec:noncritical}, we will 
briefly explain the construction of
the solitonic operators \cite{Hanada:2004im}
in the string field theory of noncritical
strings \cite{Ishibashi:1993pc}. In section \ref{sec:ospsft},
we will review the $OSp$  invariant
string field theory \cite{Siegel:1984ap}\cite{Kugo:1987rq}.
In section \ref{sec:boundary}, we will define the boundary states 
and the creation and annihilation operators for
the strings whose wave functions are
proportional to such states,
in the $OSp$ invariant string field theory.
In section \ref{sec:solitonop}, 
we will construct solitonic operators using the operators 
defined in section \ref{sec:boundary},
following the construction in the noncritical case. 
Regarding them as the creation operators for D-branes, we can get 
BRST invariant states in which D-branes are excited. 
We compute the 
vacuum amplitude and find that the operators we construct create two 
D-branes or ghost D-branes. Section \ref{sec:discussions}
will be devoted to discussions.
In the appendices, we present the details of the calculations 
to derive the BRST transformations for component fields
of the string field which are proportional to the boundary states.

\section{D-branes in Noncritical String Field Theory}\label{sec:noncritical}
Noncritical strings can be described by the string field theory 
constructed in \cite{Ishibashi:1993pc} and its generalizations. 
For simplicity, let us 
consider the $c=0$ case.\footnote{In this paper, we will follow the 
conventions of \cite{Hanada:2004im} which are different from those of 
\cite{Ishibashi:1993pc}. }
In this case, the only reparametrization 
invariant quantity which can specify the state of a closed string 
is its length $l$. Therefore we define the creation and annihilation 
operators $\psi^\dagger (l), \psi (l)$ of the string with length $l$ 
which satisfy 
\begin{equation}
[\psi (l),\psi^\dagger (l^\prime )]
=
\delta (l-l^\prime )~.
\label{noncriticalcanonical}
\end{equation}
The correlation functions can be calculated using the stochastic 
Hamiltonian 
\begin{eqnarray}
H
&=&
\int_0^\infty dl_1\int_0^\infty dl_2
(l_1+l_2)\psi^\dagger (l_1+l_2)\psi (l_1)\psi (l_2)
\nonumber
\\
& &
+g_s^2
\int_0^\infty dl_1\int_0^\infty dl_2
l_1l_2\psi^\dagger (l_1)\psi^\dagger (l_2)\psi (l_1+l_2)
\nonumber
\\
& &
+\int_0^\infty dl
\rho (l)\psi^\dagger (l)
\nonumber
\\
&=&
\int^{\infty}_{0} dl\psi^\dagger (l)(lT(l)+\rho (l))~,
\label{eq:stochastic}
\end{eqnarray}
where
\begin{eqnarray}
T(l)
&=&
\int_0^l dl^\prime \psi (l^\prime )\psi (l-l^\prime )
+g_{s}^{2}
\int_0^\infty dl^\prime l^\prime
             \psi^\dagger (l^\prime )\psi (l+l^\prime )~, 
\nonumber
\\
\rho (l)
&=&
3\delta^{\prime\prime}(l)-\frac{3}{4}\mu\delta (l)~.
\label{noncriticalhamiltonian}
\end{eqnarray}
The processes which the first two terms in the Hamiltonian represent 
are exactly the joining-splitting interactions. The third term 
corresponds to a tadpole term in which only strings with 
vanishing length are involved. Here $g_s$ denotes the string coupling 
constant and $\mu$ denotes the cosmological constant on the 
worldsheet. 

In this formulation, the Virasoro constraint for $c=0$ string theory
can be written as 
\begin{equation}
T(l)|\Psi \rangle =0~,
\label{noncriticalvirasoro}
\end{equation}
where $|\Psi \rangle$ is a state which can be expressed by using the 
correlation functions. 
Solitonic operators corresponding to D-branes can be constructed as 
operators which commute with $T(l)$ \cite{Fukuma:1996hj}. If such 
operators exist, by acting them on $|\Psi \rangle$ which is a solution 
of eq.(\ref{noncriticalvirasoro}), one can generate other solutions. 

{}From the commutation relations
\begin{eqnarray}
{}\left[\,
   \frac{1}{g_s^2}\int^{\infty}_{0} dl^\prime
      l^\prime\epsilon (l^\prime )T(l^\prime )
\, , \,
   \frac{1}{g_sl}\psi (l)
\,\right]
&=&
{}-\frac{1}{g_s}\int_0^\infty dl^\prime
      l^\prime\epsilon (l^\prime )\psi (l+l^\prime )~,
\nonumber
\\
{}\left[\,
\frac{1}{g_s^2}\int^{\infty}_{0} dl^\prime
    l^\prime\epsilon (l^\prime )T(l^\prime )
\, ,\,
   g_s\psi^\dagger (l)
\, \right]
&=&
g_s\int_0^l dl^\prime l^\prime (l-l^{\prime})
    \epsilon (l^\prime )\psi^\dagger (l-l^\prime )
\nonumber
\\
& &
{}+\frac{2}{g_s}\int_0^\infty dl^\prime 
      (l+l^\prime )\epsilon (l+l^\prime )\psi (l^\prime )~,
\label{noncriticalcommutation}
\end{eqnarray}
it is straightforward to show that 
\begin{equation}
\mathcal{V}(\zeta )
\equiv
\exp \left(g_s\int_0^\infty dle^{-\zeta l}\psi^\dagger (l)\right)
\exp \left(-\frac{2}{g_s}\int_0^\infty \frac{dl}{l}e^{\zeta l}\psi (l)
     \right)
\end{equation}
satisfies
\begin{equation}
\left[\,
\frac{1}{g_s^2}\int^{\infty}_{0}
  dl^\prime l^\prime\epsilon (l^\prime )T(l^\prime )
\,,\,
\mathcal{V}(\zeta )
\right]
=
\partial_\zeta 
 \left(\partial_\zeta \tilde{\epsilon} (\zeta )
        \mathcal{V}(\zeta )\right)~,
\label{noncritical1}
\end{equation}
where
\begin{equation}
\tilde{\epsilon} (\zeta )
=
\int_0^\infty dle^{-\zeta l}\epsilon (l)~.
\end{equation}
Therefore 
$\int d\zeta \mathcal{V}(\zeta )$
commutes with $T(l)$ if the limits of the integral are chosen 
appropriately. 
In perturbative calculations, the integration over $\zeta$ is done 
by the saddle point method and we do not have to specify these limits. 
This operator can be identified with the creation operator of
the ZZ-brane. 
$\psi (l)$ in the exponent in $\mathcal{V}(\zeta )$ has the effect of 
generating boundaries on the worldsheet with exactly the same weight
as that for the boundary state of the ZZ-brane.
Moreover one can see that the solitonic operator increases the number of
the eigenvalues of the matrix for the matrix model, by one. 

The calculations above are quite analogous to the ones
in 2D free boson theory. 
$\psi$, $\psi^\dagger$ and $T(l)$ can be compared to
the oscillator modes of the 
boson and its energy-momentum tensor respectively.
$\mathcal{V}(\zeta )$ should 
correspond to the vertex operator with conformal weight $1$. 
The condition that the right hand side of eq.(\ref{noncritical1})
be a total derivative fixes the overall factor in the exponent
of $\mathcal{V}(\zeta )$. 
Actually, from this condition alone, there exists another choice for 
$\mathcal{V}(\zeta )$ which is 
\begin{equation}
\exp \left(-g_s\int_0^\infty dle^{-\zeta l}\psi^\dagger (l)\right)
\exp \left(\frac{2}{g_s}\int_0^\infty \frac{dl}{l}e^{\zeta l}\psi (l)
     \right)~.
\end{equation}
This operator should correspond to the ghost D-brane. 

What we would like to do in this paper is to generalize the above 
construction to critical strings. Since the solitonic operator 
$\mathcal{V}(\zeta )$ generates boundaries on the worldsheet,
we should use the creation and annihilation operators of
the critical strings
whose wave functions are proportional to the 
boundary states,
in place of $\psi^\dagger$ and $\psi$ in the above construction. 
In Ref.\cite{Kishimoto:2003ru},
it was shown that the boundary states $|B\rangle$ satisfy 
the idempotency equation 
\begin{equation}
|B\rangle *|B\rangle \propto |B\rangle ,
\end{equation}
where $*$ denotes the product corresponding to a light-cone gauge type
three string vertex.
This equation implies that in the joining-splitting interaction for 
the strings whose wave function is proportional to the boundary states, 
what matter are only their lengths. 
Therefore the 
three string vertex for such states is essentially the same as the one in 
the Hamiltonian (\ref{eq:stochastic}) for noncritical strings. 
Hence it seems likely that the above construction
works also for some string field 
theory of critical strings.

\section{$OSp$ Invariant String Field Theory}\label{sec:ospsft}
The string field theory we consider in this paper is the $OSp$ invariant 
string field theory \cite{Siegel:1984ap}\cite{Kugo:1987rq}.
In order to fix the notations used in this paper,
let us recapitulate the formulation of this theory.

\subsubsection*{$OSp$ extension}

Siegel's procedure \cite{Siegel:1984ap} for covariantizing
the light-cone gauge string field
theory \cite{Kaku:1974zz}\cite{Mandelstam:1973jk}%
\cite{Cremmer:1974ej}
is to replace the $O(24)$ transverse vector $X^{i}$
by the $OSp(25,1|2)$ vector
$X^{M}=\left( \sqrt{\frac{2}{\palpha}}X^\mu ,C,\bar{C}\right)$,
where $X^{\mu}=(X^{i},X^{+},X^{-})$ are Grassmann even
and the ghost fields $C$ and $\bar{C}$ are Grassmann odd.
The metric of the $OSp(25,1|2)$ vector space is
\begin{equation}
\eta_{MN}
=
\begin{array}{c}
\\
\\
\\
\\
\mbox{\scriptsize $C$}\\
\mbox{\scriptsize $\bar{C}$}\\
\end{array}
\begin{array}{r}
\mbox{\scriptsize$C$}~~\mbox{\scriptsize$\bar{C}$}\hspace{6mm}\\
\left(
\begin{array}{ccc|cc}
 &             & & &  \\
 &\eta_{\mu\nu}& & &  \\
 &             & & &  \\\hline
 &             & &0&-i\\
 &             & &i&0 \\
\end{array}
\right)
\end{array}
 =\eta^{MN}~.
\end{equation}
The Euclidean action is
\begin{equation}
 S = \frac{1}{8\pi} \int d\tau d\sigma
     \partial_aX^M\partial^aX^{N} \eta_{MN}~,
\end{equation}
where $(\tau,\sigma)$ denote the coordinates on the
cylinder worldsheet.
One has the mode expansion
\begin{eqnarray}
X^\mu (\tau ,\sigma )
&=&  x^\mu - \palpha ip^\mu \tau 
    +i \sqrt{\frac{\palpha}{2}}\sum_{n\neq 0}\frac{1}{n}
        \left(\alpha_n^\mu e^{-n(\tau +i\sigma )}
            +\tilde{\alpha}_n^\mu e^{-n(\tau -i\sigma )}\right)~,
\nonumber \\
C(\tau ,\sigma ) &=& C_0+2i\pi_0\tau 
       -i\sum_{n\neq 0}\frac{1}{n} 
          \left(\gamma_n e^{-n(\tau +i\sigma )}
             +\tilde{\gamma}_ne^{-n(\tau -i\sigma )}\right)~,
\nonumber\\
\bar{C}(\tau ,\sigma )
&=&  \bar{C}_0-2i\bar{\pi}_0\tau 
     +i\sum_{n\neq 0}\frac{1}{n}
        \left(\bar{\gamma}_n e^{-n(\tau +i\sigma )}
           +\tilde{\bar{\gamma}}_ne^{-n(\tau -i\sigma )}\right)~.
\end{eqnarray}
The nonvanishing canonical commutation relations are 
\begin{eqnarray}
&& [x^\mu ,p^\nu ] = i\eta^{\mu\nu}~, \quad
  [ \alpha_n^\mu ,\alpha_m^\nu ] = n\eta^{\mu\nu}\delta_{n+m,0}~,
  \quad
  [\tilde{\alpha}_n^\mu ,\tilde{\alpha}_m^\nu ]
    = n\eta^{\mu\nu}\delta_{n+m,0}~,
\nonumber \\
&&\{ C_0,\bar{\pi}_0\} = 1~,
\quad \{ \bar{C}_0,\pi_0\} = 1~,
\quad \{ \gamma_n,\bar{\gamma}_m\} = in\delta_{n+m,0}~,
\quad \{ \tilde{\gamma}_n,\tilde{\bar{\gamma}}_m\}
        =  in\delta_{n+m,0}
\end{eqnarray}
for $n\neq 0$.
We also use 
\begin{equation}
\alpha^{\mu}_{0}=\tilde{\alpha}^{\mu}_{0}
  =\sqrt{\frac{\palpha}{2}} p^{\mu}~,
\quad 
\gamma_0=\tilde{\gamma}_0 \equiv \pi_0~,
\quad
\bar{\gamma}_0=\tilde{\bar{\gamma}}_0 
  \equiv \bar{\pi}_0~.
\end{equation}

The Hilbert space for the string consists of the Fock space
of the oscillators and the wave function for the zero modes.
We take the wave function to be a function of 
$p^\mu , \alpha , \pi_0, \bar{\pi}_0$,
i.e.\ we take the momentum representation for the zero modes.
Here $\alpha$ denotes the string length,
which is a variable characteristic of
the string field theories with the joining-splitting
interactions.
In the momentum representation,
the vacuum state $|0\rangle$ in the first quantization
is defined by
\begin{eqnarray}
&& \alpha^{\mu}_{n} |0\rangle=\tilde{\alpha}^{\mu}_{n}|0\rangle
  =0~, \quad
\gamma_{n}|0\rangle = \tilde{\gamma}_{n}|0\rangle
    =0~, \quad
   \bar{\gamma}_{n}|0\rangle=\tilde{\bar{\gamma}}_{n}|0\rangle
   =0
   \qquad \mbox{for $n>0$}~, \nonumber\\
&&
x^{\mu} |0\rangle
 =i\frac{\partial}{\partial p_{\mu}} |0\rangle=0~,
\quad
C_{0}|0\rangle
  =\frac{\partial}{\partial \bar{\pi}_{0}} |0\rangle=0~,
\quad
\bar{C}_{0}|0\rangle
  =\frac{\partial}{\partial \pi_{0}}|0\rangle
  =0~,\nonumber\\
&&
\frac{\partial}{\partial \alpha} |0\rangle =0~.
\end{eqnarray}
The integration measure for the zero modes 
of the $r$-th string is written as 
\begin{equation}
dr
\equiv
 (2\pi)^{-27} \alpha_rd\alpha_r
 d^{26}p_r \, i  d\bar{\pi}_0^{(r)} d\pi_0^{(r)}~.
 \label{eq:zeromodemeasure}
\end{equation}

The BRST charge is defined
\cite{Siegel:1986zi}\cite{Bengtsson:1986yj} as 
\begin{eqnarray}
 Q_{\mathrm{B}}
 &=& \frac{C_0}{2\alpha}(L_0+\tilde{L}_{0} - 2)
   -i\pi_0\frac{\partial}{\partial \alpha}
  \nonumber\\
&& 
    {} +\frac{i}{\alpha}\sum_{n=1}^{\infty}
      \left(
        \frac{\gamma_{-n}L_n-L_{-n}\gamma_n}{n}
          +\frac{\tilde{\gamma}_{-n}\tilde{L}_n
                 -\tilde{L}_{-n}\tilde{\gamma}_n}{n}\right)~,
  \label{eq:brst-charge}
\end{eqnarray}
where $L_{n}$ and $\tilde{L}_{n}$ $(n\in\mathbb{Z})$ are
the Virasoro generators defined as
\begin{eqnarray}
L_n
&\equiv& \sum_m
     \maru \left( \frac{1}{2}\alpha_{n+m}^\mu\alpha_{-m,\mu}
     +i\gamma_{n+m}\bar{\gamma}_{-m}\right)\maru~,
  \nonumber \\
\tilde{L}_n
&\equiv&
       \sum_m
      \maru \left(\frac{1}{2}\tilde{\alpha}_{n+m}^\mu
              \tilde{\alpha}_{-m,\mu}
         +i\tilde{\gamma}_{n+m}\tilde{\bar{\gamma}}_{-m}\right)
         \maru~.
\end{eqnarray}
Here $\maru\ \maru$ means the normal ordering of the oscillators
in which the non-negative modes should be moved to the right
of the negative modes. 
The BRST charge (\ref{eq:brst-charge}) is nilpotent:
$(Q_{\mathrm{B}})^2=0$.

%
%
%

\subsubsection*{The reflector}
The reflector is defined as
\begin{equation}
\left\langle R(1,2)\right|
= \delta_{\mathrm{LC}}(1,2)
\; {}_{12}\!\langle 0|
 \, e^{E(1,2)}\, \frac{1}{\alpha_1}~,
\end{equation}
where
\begin{eqnarray}
{}_{12}\! \langle 0| 
&=& {}_{1}\!\langle 0| {}_{2}\!\langle 0|~,
 \nonumber\\
E(1,2)
 &=&
  -\sum_{n=1}^\infty\frac{1}{n}
        \left(\alpha_{n}^{M(1)} \alpha_{n}^{N(2)}
        +\tilde{\alpha}_{n}^{M (1)} \tilde{\alpha}_{n}^{N(2)}
        \right)\eta_{MN}~,
\nonumber\\
\delta_{\mathrm{LC}} (1,2)
&=&
 i (2\pi )^{27}\delta (\alpha_1+\alpha_2)
\delta^{26}(p_1+p_2)
(\bar{\pi}_0^{(1)}+\bar{\pi}_0^{(2)})
(\pi_0^{(1)}+\pi_0^{(2)})~,
\end{eqnarray}
with
\begin{equation}
\alpha^{M}_{n}
  =\left(\alpha^{\mu}_{n},-\gamma_{n},\bar{\gamma}_{n}\right)~,
 \quad
   \tilde{\alpha}^{M}_{n}
   =\left(\tilde{\alpha}^{\mu}_{n},-\tilde{\gamma}_{n},
          \tilde{\bar{\gamma}}_{n}\right)~.
\end{equation}
We also introduce
\begin{equation}
|R(1,2)\rangle
=
\delta_{\mathrm{LC}} (1,2)\frac{1}{\alpha_1}
e^{E^\dagger (1,2)}|0\rangle_{12}~.
\end{equation}

The reflector $\langle R(1,2)|$ satisfies
\begin{eqnarray}
\left\langle R(1,2)\right|
  \left( \alpha_{1}+\alpha_{2}\right)=0~,
 &&
  \left\langle R(1,2)\right|
    \left( x^{\mu(1)}-x^{\mu(2)} \right)=0~,
   \nonumber\\
\left\langle R(1,2)\right|
     \left(C_{0}^{(1)}-C_{0}^{(2)}\right)=0~,
  &&
  \left\langle R(1,2)\right|
     \left(\bar{C}_{0}^{(1)}-\bar{C}_{0}^{(2)} \right)=0~,
     \nonumber\\
\left\langle R(1,2)\right|
      \left(\alpha^{M(1)}_{n} + \alpha^{M(2)}_{-n}\right)
      =0~,
  &&
   \left\langle R(1,2)\right|
    \left(\tilde{\alpha}^{M(1)}_{n}
           + \tilde{\alpha}^{M(2)}_{-n}\right)
      =0 \quad \mbox{for $\forall n \in \mathbb{Z}$}~.
\label{eq:reflection1}
\end{eqnarray}
This yields
\begin{eqnarray}
&& \left\langle R(1,2)\right|
      \left( L^{(1)}_{n}-L^{(2)}_{-n} \right)=0~,
\quad
   \left\langle R(1,2)\right|
      \left( \tilde{L}^{(1)}_{n}
            -\tilde{L}^{(2)}_{-n} \right)=0
   \quad \mbox{for $\forall n \in \mathbb{Z}$}
   \nonumber \\
&& \left\langle R(1,2)\right|
   \left(Q_B^{(1)}+Q_B^{(2)}\right)=0~.
\label{eq:reflection2}
\end{eqnarray}
$\left|R(1,2)\right\rangle$ satisfies similar identities. 


The BPZ conjugate $\langle \Phi |$ of $|\Phi\rangle$
is defined as 
\begin{equation}
{}_{2} \! \langle \Phi |
=
\int d1\, \langle R(1,2)|\Phi \rangle_1~.
\label{eq:BPZ}
\end{equation}

{}From the definitions, we have
\begin{equation}
\int d1d2\,
  \langle R(1,2)|\Phi\rangle_1|\Psi\rangle_2
=
-(-1)^{|\Phi ||\Psi |}
\int d1d2\, \langle R(1,2)|\Psi\rangle_1|\Phi\rangle_2~,
\end{equation}
and
\begin{equation}
\int d1
\, {}_{1}\!\langle \Phi |R(1,2)\rangle
=|\Phi\rangle_2~,
\end{equation}
where $(-1)^{|\Phi|}$ denotes the Grassmann parity
of the string field $\Phi$.
Thus $\langle R(1,2)|$ can be considered as the 
symplectic form for the string fields and 
$|R(1,2)\rangle$ is its inverse. 

\subsubsection*{The three string vertex}
The three string vertex is given by 
\begin{eqnarray}
\lefteqn{
\left\langle V_3(1,2,3)\right|
=
\delta_{\mathrm{LC}} (1,2,3)
   \; {}_{123}\!\langle 0|e^{E (1,2,3)}
    C(\rho_I)
    \mathcal{P}_{123}
    \frac{|\mu (1,2,3)|^2}{\alpha_1\alpha_2\alpha_3}
    }\nonumber \\
&&= i \delta (1,2,3)
  \; {}_{123}\!\langle 0|e^{E (1,2,3)}
    \left( \sum_{r=1}^3\bar{\pi}_0^{(r)} \right)
    \left( \sum_{s=1}^3\pi_0^{(s)} \right)
    C(\rho_I)
{\cal P}_{123}
\frac{|\mu (1,2,3)|^2}{\alpha_1\alpha_2\alpha_3}~,
\label{eq:V3}
\end{eqnarray}
where $\rho_{I}$ denotes the interaction point and
\begin{eqnarray}
{}_{123}\!\langle 0|
 &=& {}_{1}\!\langle 0 |\,{}_{2}\!\langle 0|\,{}_{3}\!\langle 0|~,
\nonumber\\
\mathcal{P}_{123}&=&\mathcal{P}_{1}\mathcal{P}_{2}\mathcal{P}_{3}~,
\quad \mathcal{P}_{r}=\int^{2\pi}_{0} \frac{d\theta}{2\pi}
      e^{i\theta \left(L^{(r)}_{0}-\tilde{L}^{(r)}_{0}\right)}~,
\nonumber\\
\delta_{\mathrm{LC}} (1,2,3)
&=&
 i  (2\pi )^{27}
\, \delta^{26} \left( \sum_{r=1}^3 p_r \right)
\, \delta \left( \sum_{s=1}^3 \alpha_s \right)
\, \left(\sum_{r^\prime =1}^3\bar{\pi}_0^{(r^\prime )}\right)
\left( \sum_{s^\prime =1}^3\pi_0^{(s^\prime )} \right)~,
\nonumber\\
\delta (1,2,3)
&=&
(2\pi )^{27}
\, \delta^{26} \left( \sum_{r=1}^3 p_r \right)
\, \delta \left(\sum_{s=1}^3 \alpha_s \right)~,
\nonumber\\
E(1,2,3)
&=&
\sum_{n,m\geq 0}\sum_{r,s}
\bar{N}_{nm}^{rs}
 \left( \frac{1}{2}\alpha_n^{\mu (r)} \alpha_{m\mu}^{\ (s)}
      +i\gamma_n^{(r)}\bar{\gamma}_m^{(s)}
      +\frac{1}{2}\tilde{\alpha}_n^{\mu (r)}
           \tilde{\alpha}_{m\mu}^{\ (s)}
      +i\tilde{\gamma}_n^{(r)}\tilde{\bar{\gamma}}_m^{(s)}
  \right)~.
\nonumber \\
\mu(1,2,3) & = &
  \exp\left(-\hat{\tau}_{0} \sum_{r=1}^{3}\frac{1}{\alpha_{r}} 
      \right)~,
    \quad \hat{\tau}_{0}
    =\sum_{r=1}^{3} \alpha_{r} \ln \left|\alpha_{r}\right|~.
\end{eqnarray}
Here $\bar{N}^{rs}_{nm}$ denote the Neumann coefficients
associated with the joining-splitting type of three string
interaction \cite{Kaku:1974zz}%
\cite{Mandelstam:1973jk}\cite{Cremmer:1974ej}.\footnote{Although
the Neumann coefficients in
the anti-holomorphic sector are complex conjugate to
those in the holomorphic sector in general,
one may choose the Neumann coefficients for the three string
vertex to be real because of the $SL(2,\mathbb{C})$ invariance
on the worldsheet.}


By using the three string vertex (\ref{eq:V3}),
the $\ast$-product $\Phi \ast \Psi$
of two arbitrary closed string fields
$\Phi$ and $\Psi$ is defined by
\begin{equation}
\left|\Phi *\Psi \right\rangle_4
=
\int d1d2d3\, \left\langle V_3(1,2,3)\left|
     \Phi\right\rangle_1 \right.
     \left|\Psi \right\rangle_2
     \left| R(3,4) \right\rangle~.
 \label{eq:star-prod}
\end{equation}
The $\ast$-product has following properties,
\begin{eqnarray}
&&Q_{\mathrm{B}}\left( \Phi \ast \Psi \right)
  = \left( Q_{\mathrm{B}} \Phi \right) \ast \Psi
     + (-1)^{|\Phi|}\Phi \ast \left(Q_{\mathrm{B}} \Psi\right)~,
\nonumber\\
&& \left(\Phi_{1} \ast \Phi_{2}\right) \ast \Phi_{3}
  +(-1)^{|\Phi_{1}|\left(|\Phi_{2}|+|\Phi_{3}|\right)}
    \left( \Phi_{2} \ast \Phi_{3} \right) \ast \Phi_{1}
 \nonumber\\
 && \hspace{6.5em}{}  +
   (-1)^{|\Phi_{3}|\left(|\Phi_{1}|+|\Phi_{2}|\right)}
    \left( \Phi_{3} \ast \Phi_{1} \right) \ast \Phi_{2}
 =0~.
 \label{eq:prop-ast}
\end{eqnarray}
The first equation is equivalent to
\begin{equation}
\left\langle V_{3} (1,2,3) \right|
  \sum_{r=1}^{3} Q_{\mathrm{B}}^{(r)} =0~,
\end{equation}
and the second one is known as the Jacobi identity.

\subsubsection*{String field action}
The action of the $OSp$ invariant string field theory is directly
given by the $OSp$ extension from that of the light-cone gauge
string field theory.
This takes the form
\begin{eqnarray}
\lefteqn{
S = \int dt  \left[
  \frac{1}{2}
   \int d1d2\, \left\langle R(1,2) \left|\Phi\right\rangle_{1}
               \right.
       \left( i\frac{\partial}{\partial t}
                  -\frac{L_0^{(2)}+\tilde{L}_0^{(2)}-2}{\alpha_2}
       \right)  \left|\Phi\right\rangle_{2}
   \right.} \hspace{3.5em}\nonumber\\
&& \left.
+ \frac{2g}{3}
\int d1d2d3 \, \left\langle V_3(1,2,3)\right|
    \left(\sum_{r=1}^{3} \bar{\pi}_{0}^{(r)}\right)
|\Phi\rangle_1|\Phi\rangle_2|\Phi\rangle_3
\right]~,
\label{eq:actionOSp}
\end{eqnarray}
where $t$ denotes the proper time.
The string field $\Phi$ is taken to be Grassmann even
and subject to the level matching condition
$\mathcal{P}\Phi=\Phi$.
Note that in the interaction term
the three string vertex $\left\langle V_{3}(1,2,3) \right|$
is multiplied by the factor $\sum_{r=1}^{3}\bar{\pi}^{(r)}_{0}$.
This manipulation removes $C(\rho_{I})$ from the vertex
$\left\langle V_{3}(1,2,3) \right|$, i.e.
\begin{eqnarray}
\left\langle V_{3}^{0}(1,2,3) \right|
 \equiv  \left\langle V_3(1,2,3)\right|
    \left(\sum_{r=1}^{3} \bar{\pi}_{0}^{(r)}\right)
 = \delta_{\mathrm{LC}} (1,2,3)
   \; {}_{123}\!\langle 0|e^{E (1,2,3)}
    \mathcal{P}_{123}
    \frac{|\mu (1,2,3)|^2}{\alpha_1\alpha_2\alpha_3}~.
\label{V30}
\end{eqnarray}

The action (\ref{eq:actionOSp}) is
invariant under the BRST transformation
\begin{equation}
\delta_{\mathrm{B}} \Phi
 =Q_{\mathrm{B}} \Phi +g\Phi *\Phi~,
\label{BRS}
\end{equation}
where the $\ast$-product is defined in eq.(\ref{eq:star-prod}).
The nilpotency of the BRST transformation (\ref{BRS})
is assured by the nilpotency of $Q_{\mathrm{B}}$ and
eqs.(\ref{eq:prop-ast}).
One can readily show that the action (\ref{eq:actionOSp})
is invariant under the BRST transformation (\ref{BRS})
by using the nilpotency of the BRST transformation (\ref{BRS})
and the fact that the action (\ref{eq:actionOSp})
can be expressed as
\begin{equation}
S= \int dt\left[ \frac{1}{2} \int d1 d2 \langle R(1,2)|
     \Phi\rangle_{1} i\frac{\partial}{\partial t}
     |\Phi\rangle_{2}
     + \delta_{\mathrm{B}}\left(
      \int d1 d2 \langle R(1,2) |\Phi\rangle_{1}
         \bar{\pi}_{0}^{(2)} |\Phi\rangle_{2} \right) \right]~.
   \label{eq:BRSTexact}
\end{equation}

In this string field theory, $C$ and $\bar{C}$ play the role 
of the $b,c$ ghost in the usual theory. Indeed with the following 
identifications 
\begin{eqnarray}
&& \gamma_n = in\alpha c_n~,
   \quad  \tilde{\gamma}_n = in\alpha\tilde{c}_n~;
   \quad  \bar{\gamma}_n = \frac{1}{\alpha}b_n~,
   \quad \tilde{\bar{\gamma}}_n = \frac{1}{\alpha}\tilde{b}_n~,
  \nonumber \\
 && C_0 = 2\alpha c_0^+~,
    \quad \bar{\pi}_0 = \frac{1}{2\alpha}b_0^+~,
\label{ghostidentification}
\end{eqnarray}
with $n\neq 0$, $Q_B$ becomes almost the same as the usual 
first-quantized BRST operator. 
Perturbative calculations can be done in a way similar to
the one for the light-cone gauge string field theory. 
In the canonical quantization, we should think of the 
components of $|\Phi \rangle$ with positive $\alpha$
as annihilation operators and those with 
negative $\alpha$ as creation operators. 
The prescription for how to treat the
physical on-shell states was given by
Ref.\cite{Kugo:1987rq}\footnote{While the author
of Ref.\cite{Kugo:1987rq}
gives the prescription in the context of the gauge invariant 
covariantized light-cone string field theory,
his prescription is applicable to
the $OSp$ invariant string field theory as well.},
and a proof was given to the fact that the S-matrix elements calculated 
using this theory coincide with those of the light-cone 
gauge string field theory. 

Before concluding this section, one comment is in order. 
In Refs.\cite{Neveu:1986cu}\cite{Uehara:1987qz}
\cite{Kugo:1987rq}\cite{Kawano:1992dp}, gauge invariant actions
were proposed and it was shown that the $OSp$ invariant theory 
can be obtained from them after gauge fixing. 
Unfortunately the BRST transformations which originate from these 
covariantized light-cone string field theories coincide with 
eq.(\ref{BRS}) only for on-shell states. 
In this paper, we should deal with the boundary states which are 
off-shell, and consider eq.(\ref{BRS}) as the BRST transformation. 
The origin of this BRST transformation eq.(\ref{BRS}) may be understood by 
considering this system in terms of the BFV formalism. In principle, 
looking at the BRST transformation itself, 
one can read off the constraints from which the BRST transformation 
is constructed.

\section{Boundary State and String Field}\label{sec:boundary}
In order to construct solitonic operators in the way mentioned at 
the end of section \ref{sec:noncritical},
we should study the boundary states 
in the $OSp$ invariant theory and identify the creation and annihilation 
operators corresponding to such states. 
A problem is that the boundary states are not normalizable. 
We will introduce a BRST invariant regularization and define normalizable 
states proportional to the boundary states. 

In what follows, we consider the toroidally compactified
space-time characterized by $X^{\mu} \simeq X^{\mu}+2\pi R^{\mu}$
$(\mu=0,1,\ldots,25)$
to regularize the infrared divergence.
In this situation, the zero modes of the matter sector
are modified because of the momentum quantization and
the windings. 
We briefly summarize the notations for the zero-mode part
of the toroidally compactified matter sector.

The zero-mode part of $X^{\mu}(\tau,\sigma)$
takes the form
\begin{equation}
   X^{\mu}(\tau,\sigma) \Big|_{\mbox{\scriptsize zero-mode}}
    =x_{0}^{\mu} +\palpha \left( -ip^{\mu}\tau + q^{\mu}\sigma \right)
       \nonumber \\
    = x^{\mu}_{L}+x^{\mu}_{R} -i\frac{\palpha}{2} 
           \left(p^{\mu}_{L} \ln w + p^{\mu}_{R} \ln \bar{w} \right)~,
\end{equation}
where $w$ and $\bar{w}$ are complex coordinates on
the cylinder worldsheet defined as
$w=e^{\tau+i\sigma}$ and $\bar{w}=e^{\tau-i\sigma}$.
The center-of-mass momentum $p^{\mu}$ is quantized
and $q^{\mu}$ is related to the winding number
$\ww^{\mu}$ as follows:
\begin{equation}
p^{\mu}=\frac{\nn^{\mu}}{R^{\mu}}~,
\quad q^{\mu}= \frac{R^{\mu}\ww^{\mu}}{\palpha}~,
\qquad \nn^{\mu},\ \ww^{\mu} \in \mathbb{Z}.
\label{eq:wind}
\end{equation}
$x^{\mu}_{L,R}$ and $p^{\mu}_{L,R}$ are defined as
\begin{eqnarray}
x^{\mu}_{0}=x^{\mu}_{L}+x^{\mu}_{R}~,&&
p^{\mu}_{L} =p+q
=\sqrt{\frac{2}{\palpha}} \alpha^{\mu}_{0}
   =\left( \frac{\nn^{\mu}}{R^{\mu}}+\frac{R^{\mu}\ww^{\mu}}{\palpha}
   \right)~,\nonumber\\
&&
p^{\mu}_{R} = p-q
=\sqrt{\frac{2}{\palpha}} \tilde{\alpha}^{\mu}_{0}
 =\left( \frac{\nn^{\mu}}{R^{\mu}}-\frac{R^{\mu}\ww^{\mu}}{\palpha}
   \right)~.
\end{eqnarray}
They obey the canonical commutation relations:
$\left[ x^{\mu}_{L}\, , \,p^{\nu}_{L}\right]=
\left[ x^{\mu}_{R}\, , \, p^{\nu}_{R} \right]=i \eta^{\mu\nu}$,
otherwise vanishing.
Thus the zero-mode sector consists of
two canonical pairs.
For later use, we introduce a new variable
$y_{0}^{\mu} \equiv x^{\mu}_{L}-x_{R}^{\mu}$
and
we choose the basis of the zero-mode phase space to be
 $\{x^{\mu}_{0},y^{\mu}_{0};p^{\mu},q^{\mu}\}$.
They satisfy $\left[x^{\mu}_{0}\, ,\, p^{\nu}\right]
=\left[y^{\mu}_{0}\, , \, q^{\nu}\right]=i\eta^{\mu\nu}$.
Because of the quantization (\ref{eq:wind}) of $q^{\mu}$,
the range in which $y^{\mu}_{0}$ (conjugate to $q^{\mu}$)
varies is finite
as well as that of $x^{\mu}_{0}$:
\begin{equation}
0\leq x^{\mu}_{0} \leq 2\pi R^{\mu}~,
\quad 0\leq y^{\mu}_{0} \leq \frac{2\pi \palpha}{R^{\mu}}~.
\label{eq:range}
\end{equation}
Let $\left| x^{\mu} \right\rangle$
and $\left| y^{\mu} \right\rangle$
be the eigenstates of the operators $x^{\mu}_{0}$ and $y^{\mu}_{0}$
with eigenvalues $x^{\mu}$ and $y^{\mu}$.
Let $\left|\nn^{\mu}\right\rangle$ and
$\left| \ww^{\mu} \right\rangle$ be the eigenstates
of the operators $p^{\mu}$ and $q^{\mu}$
with eigenvalues $p^{\mu}=\frac{\nn^{\mu}}{R^{\mu}}$
and $q^{\mu}= \frac{R^{\mu}\ww^{\mu}}{\palpha}$.
We normalize these states as follows:
\begin{eqnarray}
&&\left\langle x^{\prime\mu} \left| x^{\mu} \right\rangle \right.
=\delta\left(x^{\prime\mu}-x^{\mu}\right)~,
\quad
\left\langle y^{\prime\mu} \left| y^{\mu} \right\rangle \right.
=\delta\left(y^{\prime\mu}-y^{\mu}\right)~,
\nonumber\\
&& \left\langle \nn^{\prime\mu} \left| \nn^{\mu} \right\rangle\right.
   =\delta_{\nn^{\prime\mu},\nn^{\mu}}~,
\quad
\left\langle \ww^{\prime\mu} \left| \ww^{\mu} \right\rangle\right.
   =\delta_{\ww^{\prime\mu},\ww^{\mu}}~.
   \label{eq:normes}
\end{eqnarray}
It follows from eqs.(\ref{eq:range}) and (\ref{eq:normes}) that
\begin{equation}
\left| x^{\mu} \right\rangle
 = \frac{1}{\sqrt{2\pi R^{\mu}}}
   \sum_{\nn^{\mu} \in \mathbb{Z}} e^{-i\frac{\nn^{\mu}}{R^{\mu}}}
   \left|\nn^{\mu}\right\rangle~,
\quad
\left| y^{\mu} \right\rangle
  =\sqrt{\frac{R^{\mu}}{2\pi\palpha}} \sum_{\ww^{\mu}\in\mathbb{Z}}
  e^{-i\frac{R^{\mu}\ww^{\mu}}{\palpha} y^{\mu}}
  \left| \ww^{\mu} \right\rangle~.
\end{equation}

In accordance with the modification above,
the momentum dependent parts of
the integration measure for the zero modes,
the reflector
$\left\langle R(1,2) \right|$ and the
three string vertex $\langle V_{3}(1,2,3)|$
are respectively replaced as follows:
\begin{eqnarray}
\mbox{measure}:&& \int \frac{d^{26}p}{(2\pi)^{26}}
       \longrightarrow
  \prod_{\mu=0}^{25}
  \left( \sum_{\nn^{\mu}\in\mathbb{Z}}
         \sum_{\ww^{\mu} \in \mathbb{Z}} \right)~,
   \\
\left\langle R(1,2)\right|:
 && (2\pi)^{26} \delta^{26}(p_{1}+p_{2})
       \longrightarrow
\prod_{\mu=0}^{25} \left(
   \delta_{\nn^{\mu}_{1}+\nn^{\mu}_{2},0}
   \delta_{\ww^{\mu}_{1}+\ww^{\mu}_{2},0}
      \right)~,\nonumber\\
\left\langle V_{3}(1,2,3)\right|:
 && (2\pi)^{26} \delta^{26} \left(p_{1}+p_{2}+p_{3}\right)
    \longrightarrow
     \prod_{\mu=0}^{25} \left(
   \delta_{\nn^{\mu}_{1}+\nn^{\mu}_{2}+\nn^{\mu}_{3},0}
   \delta_{\ww^{\mu}_{1}+\ww^{\mu}_{2}+\ww^{\mu}_{3},0}
      \right) 
      e^{-i\pi(\nn_{3}\cdot \ww_{2}-\nn_{1}\cdot\ww_{1})}~.
   \nonumber
\end{eqnarray}
In the last equation, the cocycle factor
$e^{-i\pi(\nn_{3}\cdot \ww_{2}-\nn_{1} \cdot \ww_{1})}$
is necessary for the Jacobi identity to be satisfied 
\cite{Hata:1986mz}\cite{Maeno:1989uc}.
(See also the second paper in Ref.\cite{Kishimoto:2004jk}.)

\subsection{Boundary state}

We consider the situation in which the D$p$-branes
extend in the $x^{\mu}$ directions with
$\mu=0,1,\ldots,p$.
We refer to these directions as the Neumann directions
and  denote them by $x^{\mu}$ ($\mu \in \mathrm{N})$.
We refer to the directions transverse to the
D-branes as the Dirichlet directions
and denote them by $x^{i}$ ($i\in \mathrm{D}$).
In the first quantized approach to the closed string,
the D-brane is described by the boundary state.

The boundary state in the matter sector
$\left|B^{X}\right\rangle$ is expressed as
the direct product of those for the Neumann
and the Dirichlet sectors:
\begin{eqnarray}
\lefteqn{
\left|B^{X}\right\rangle
  = \left|B^{X}_{\mathrm{N}} \right\rangle
   \otimes \left|B^{X}_{\mathrm{D}} \right\rangle}
  \nonumber\\
&& \left|B^{X}_{\mathrm{N}} \right\rangle
 = \frac{\sqrt{V_{\mathrm{N}}}}
          {\left(8\pi^{2}\palpha\right)^{\frac{p+1}{4}}}
    \prod_{\mu\in \mathrm{N}}\left(
        e^{-\sum_{n=1}^{\infty}
           \frac{1}{n} \alpha^{\mu}_{-n} \tilde{\alpha}_{-n\mu}}
       \,\delta_{\nn^{\mu},0}
       \sum_{\ww^{\prime \mu}\in\mathbb{Z}}
           \delta_{\ww^{\mu},\ww^{\prime\mu}}
           \right) |0\rangle~, \nonumber\\
&& \left|B_{\mathrm{D}}^{X} \right\rangle
 = \frac{\left(2\pi^{2}\palpha\right)^{\frac{26-(p+1)}{4}}}
          {\sqrt{V_{\mathrm{D}}}}
   \prod_{i\in \mathrm{D}}
    \left( e^{\sum_{n=1}^{\infty} \frac{1}{n}
               \alpha^{i}_{-n} \tilde{\alpha}^{i}_{-n}}
         \sum_{\nn^{\prime i}\in\mathbb{Z}} 
             e^{-i\frac{\nn^{\prime i}}{R^{i}}x^{i}}
             \delta_{\nn^{i},\nn^{\prime i}}
             \delta_{\ww^{i},0}
    \right) |0\rangle~,
\end{eqnarray}
where $V_{\mathrm{N}}$ and $V_{\mathrm{D}}$
are respectively the volumes of the Neumann and the Dirichlet
directions,
i.e.\ $V_{\mathrm{N}}
       =\prod_{\mu\in \mathrm{N}}\left(2\pi R^{\mu}\right)$
and $V_{\mathrm{D}}=\prod_{i\in \mathrm{D}}\left(2\pi R^{i}\right)$.
We notice that the zero-mode part of
the state $\left|B^{X}_{\mathrm{N}}\right\rangle$ is
$\left\langle \nn^{\mu}\left| \nn^{\mu}=0\right\rangle\right.
   \otimes 
 \left\langle \ww^{\mu} \left| y^{\mu}=0\right\rangle\right.$
and that of the state $\left|B^{X}_{\mathrm{D}}\right\rangle$
is $\left\langle \nn^{i}\left| x^{i} \right\rangle\right.
     \otimes
     \left\langle \ww^{i} \left| \ww^{i}=0 \right \rangle\right.$,
modulo normalization constants.
In the following, we restrict ourselves to the case in which
D$p$-branes are located at $x^{i}=0$.

Let us turn to the ghost sector.
We require that the Dirichlet boundary
condition should be satisfied by the ghost fields
$C(\tau,\sigma)$ and $\bar{C}(\tau,\sigma)$ at $\tau=0$
 on the boundary state:
\begin{equation}
 C(0,\sigma)\left|B^{\mathrm{gh}}\right\rangle=0~,
 \quad
 \bar{C}(0,\sigma) \left| B^{\mathrm{gh}} \right\rangle=0~.
 \label{eq:bc1gh}
\end{equation}
In terms of the oscillation modes, these conditions read
\begin{eqnarray}
&C_{0}\left| B^{\mathrm{gh}} \right\rangle=0~,&
\bar{C}_{0}\left| B^{\mathrm{gh}} \right\rangle=0~,\nonumber\\
& \left( \gamma_{n} - \tilde{\gamma}_{-n} \right)
   \left| B^{\mathrm{gh}} \right\rangle =0~,&
\left(\bar{\gamma}_{n}-\tilde{\bar{\gamma}}_{-n} \right)
   \left| B^{\mathrm{gh}} \right\rangle =0
\label{eq:bc2gh}
\end{eqnarray}
for $\forall n\in \mathbb{Z}$.\footnote{While the $n=0$ case
of eq.(\ref{eq:bc2gh}) is not derived from eq.(\ref{eq:bc1gh}),
it holds automatically  by definition of
$(\gamma_{0},\bar{\gamma}_{0})$
and $(\tilde{\gamma}_{0},\tilde{\bar{\gamma}}_{0})$.}
They coincide with the usual boundary conditions for the $b,c$ ghosts 
assuming eq.(\ref{ghostidentification}). 
This implies that the boundary state
$\left|B^{\mathrm{gh}}\right\rangle$ is proportional to
the state
\begin{equation}
|B_{0}^{\mathrm{gh}}\rangle
 = e^{\sum_{n=1}^{\infty} \frac{i}{n}
  \left(\gamma_{-n}\tilde{\bar{\gamma}}_{-n}
         + \tilde{\gamma}_{-n}\bar{\gamma}_{-n}\right)}
  |0\rangle~.
\end{equation}

Let us define $\left|B_{0}\right\rangle$
as
\begin{equation}
\left| B_{0} \right\rangle
  =\mathcal{N} \left|B^{X}\right\rangle \otimes
               | B^{\mathrm{gh}}_{0} \rangle~,
\label{eq:B0}
\end{equation}
where $\mathcal{N}$ is an arbitrary normalization constant.
In what follows, we refer to this state as
a boundary state.
Since the string field should have $\alpha$ dependence, we need 
an $\alpha$ dependent version of the boundary state. 
Let us define $\left| B_{0} (l) \right\rangle$ as 
\begin{equation}
\left| B_{0} (l) \right\rangle
  = \left| B_{0} \right \rangle
     \delta(\alpha-l)~,
  \label{eq:Bel}
\end{equation}
where the parameter $l$ is an eigenvalue of $\alpha$,
i.e.\ %
$\alpha \left|B_{0}(l)\right\rangle
   = l \left| B_{0}(l) \right\rangle$.

\subsection{Regularization}\label{sec:regularization}

\begin{figure}[htbp]
\vspace{1.5\baselineskip}
\begin{center}
\includegraphics[width=35em,clip]{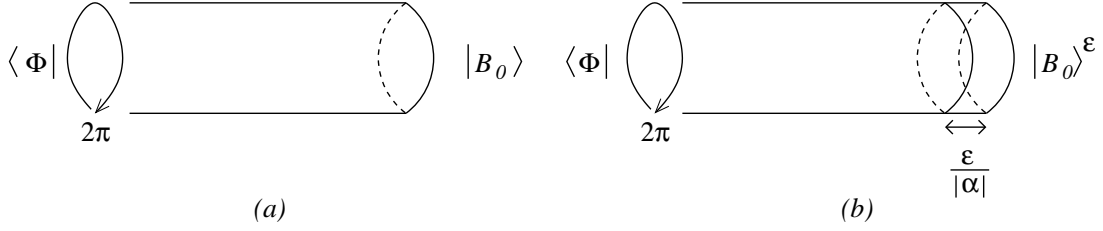}
\end{center}
\caption{(a) The inner product
   $\left\langle \Phi \left| B_{0}\right\rangle\right.$.
   (b) The inner product
   $\left\langle \Phi \left| B_{0}\right\rangle^{\epsilon}
     \right.$.}\label{fig:stab}
\end{figure}

The boundary state (\ref{eq:B0}) is not normalizable.
We need therefore regularize the divergence
of the norm of this state,
in order to treat it as a string field
in string field theory.
For this purpose
we introduce a regularized boundary state
$\left| B_{0} \right\rangle^{\epsilon}$
by attaching a stub to the state $\left| B_{0} \right\rangle$
as depicted in Fig.~\ref{fig:stab}:
\begin{equation}
  \left| B_{0} \right\rangle^{\epsilon}
    =e^{-\frac{\epsilon}{|\alpha|}\left(L_{0}+\tilde{L}_{0}-2\right)}
    \left| B_{0} \right\rangle~.
    \label{eq:regbs0}
\end{equation}
A similar regularization is necessary even for on-shell physical 
states\cite{Kugo:1987rq}. 
This is a BRST invariant regularization\footnote{
For BRST invariance, one may also use 
$\epsilon\left( L_{0}+\tilde{L}_{0}-2i\pi_0\bar{\pi}_0-2\right)$ instead of 
$\frac{\epsilon}{\alpha}\left( L_{0}+\tilde{L}_{0}-2\right)$ 
because
\[
\left\{ Q_{\mathrm{B}}\, , \, 2\epsilon \alpha\bar{\pi}_{0} \right\}
= \epsilon\left( L_{0}+\tilde{L}_{0}-2i\pi_0\bar{\pi}_0-2\right)~.
\]
This regularization however does not work for our purpose. 
See the comment at the end of this subsection. 
} 
because 
$e^{-\frac{\epsilon}{|\alpha|}\left(L_{0}+\tilde{L}_{0}-2\right)}$
commutes with the BRST charge $Q_{\mathrm{B}}$, which can be seen 
from 
\begin{equation}
\left\{ Q_{\mathrm{B}}\, , \, 2\epsilon \bar{\pi}_{0} \right\}
= \frac{\epsilon}{\alpha}\left( L_{0}+\tilde{L}_{0}-2\right)~.
\end{equation}
A subtlety seems to occur at $\alpha=0$,
because $\alpha$ appears in the form of the absolute value.
As is usual in a light-cone formalism, we will exclude the 
modes with $|\alpha |<\delta $ with some small $\delta$ from 
the spectrum. As we will see, this corresponds to an infrared 
cut-off on the worldsheet. 
We will study the theory perturbatively. Therefore we will 
keep the most dominant contributions in the limit 
$\epsilon\rightarrow 0$ at each order in $g$, in the following. 

We regularize the state (\ref{eq:Bel}) accordingly:
\begin{equation}
\left| B_{0} (l) \right\rangle^{\epsilon}
  = \left| B_{0} \right\rangle^{\epsilon}
    \delta(\alpha-l)~.
 \label{eq:regbs1}
\end{equation}
Let us consider the inner product 
\begin{equation}
\int dr\;
{}^{\epsilon}_{r}\! \left\langle B_{0}(l) \left|
   B_{0}(l') \right. \right\rangle^{\epsilon}_{r}
=\int dr ds
  \left\langle R(s,r)\left| B_{0} (l) \right. 
       \right\rangle^{\epsilon}_{s}
  \left|B_{0} (l') \right\rangle^{\epsilon}_{r}~,
\label{eq:bb-0}
\end{equation}
where $dr$ and $ds$ denote the integration measures
for the zero modes
defined in eq.(\ref{eq:zeromodemeasure}).
Eq.(\ref{eq:bb-0})
becomes the cylinder amplitude
of a closed string with a fixed circumference $2 \pi$
 propagating
through a very short proper time $\frac{2\epsilon}{|l|}$.
After the modular transformation
for the non-zero mode part and the Poisson
resummation for the zero-mode part, we obtain
\begin{eqnarray}
\int dr\;
{}^{\epsilon}_{r}\! \left\langle B_{0} (l) \left|
   B_{0} (l') \right. \right\rangle^{\epsilon}_{r}
&=& \mathcal{N}^{2}
    \, l' \delta(l+l')
    \, e^{\frac{\pi^{2}|l|}{\epsilon}}
    \, \frac{1}{\left[\prod_{m=1}^{\infty}
              \left(1-e^{-\frac{\pi^{2}|l|}{\epsilon}m}\right)
             \right]^{24}}
     \nonumber\\
 &&  \times
     \prod_{\mu \in \mathrm{N}}
     \left[ \sum_{\ww^{\mu}\in\mathbb{Z}}
      e^{-\frac{|l|}{\epsilon}
          \frac{\pi^{2}\palpha}{(R^{\mu})^{2}} \ww^{\mu 2}}
      \right]
      \times \prod_{i \in \mathrm{D}}
        \left[ \sum_{\nn^{i}\in \mathbb{Z}}
              e^{-\frac{|l|}{\epsilon}\frac{(\pi R^{i})^{2}}{\palpha}
                  \nn^{i2}} \right]~.
    \label{eq:overlap2}
\end{eqnarray}
The dominant contribution $e^{\frac{\pi^{2}|l|}{\epsilon}}$ in the limit 
$\epsilon \rightarrow 0$ 
originates 
from the propagation of the open string tachyon in the dual channel. 
We introduce the state
$\left| n(l) \right\rangle$ defined as
\begin{equation}
 \left| n(l) \right\rangle 
  = \left| B_{0} (l) \right\rangle^{\epsilon}
     e^{-\frac{\pi^{2}}{2\epsilon}|l|}~.
   \label{eq:n0-1}
\end{equation}
{}From eq.(\ref{eq:overlap2}), we find that
\begin{equation}
 \int dr {}_{r}\! \left\langle n(l) | n(l') \right\rangle_{r}
    =  \left(\mathcal{N}^{2}
            +\mathcal{O}\left(e^{-\frac{1}{\epsilon}} \right)
         \right)l'
     \, \delta(l+l')~.
\label{eq:innerprod-n}
\end{equation}
Thus the state $\left| n(l) \right\rangle$
is normalizable.

A comment is in order. Naively speaking
\begin{equation}
\int dr\;
{}_r\! \left\langle B_{0} (l) \left|
   B_{0} (l') \right. \right\rangle_{r}
=
0,
\end{equation}
because the wave function for $\left| B_{0} (l) \right\rangle$ 
lacks factors of $\pi_0$ and $\bar{\pi}_0$. However 
in eq.(\ref{eq:bb-0}), 
$e^{-\frac{\epsilon}{|\alpha|}\left(L_{0}+\tilde{L}_{0}-2\right)}$ 
provides these and we get a nonvanishing answer for the inner 
product. 

\subsection{An expansion of the string field}\label{sec:expansion}

Now let us define the creation and annihilation operators 
of the closed strings whose wave functions are proportional to 
the boundary states. 
The states
$\left\{ \left| n(l) \right\rangle,
         \left| n(-l) \right\rangle \right\}$
with $l>0$ are normalizable as stated above
and the inner products (\ref{eq:innerprod-n}) among them
are non-degenerate.
This enables us to choose a complete basis of the Hilbert space 
which consists of these states and their orthogonal complement.
Taking also the states $\bar{\pi}_{0} \left| n(l) \right\rangle  $ 
into account, we can expand $\left| \Phi \right\rangle$ as 
\begin{equation}
\left| \Phi \right\rangle
  = \int^{\infty}_{0} dl 
   \Big[ \left| n(l) \right\rangle  \phi(l)
          + \bar{\pi}_{0} \left| n(l) \right\rangle  \chi(l)
         +\left| n(-l) \right\rangle \bar{\phi}(l)
              + \bar{\pi}_{0} \left| n(-l) \right\rangle
                 \bar{\chi}(l)
             + \cdots \Big]~,
      \label{eq:sf1}
\end{equation}
where `$\cdots$' denotes the contributions from the other states. 
The wave functions $\phi (l)$, $\bar{\phi}(l)$, $\chi(l)$,
$\bar{\chi}(l)$ etc.\ %
are the fields to be quantized in the second
quantization.\footnote{The wave functions $\phi (l)$ etc.\ depend
on the proper time $t$: $\phi(t,l)$. We suppress the proper time in
the arguments for simplicity.} 
$\phi (l)$ and $\bar{\phi}(l)$ can be considered as the annihilation and 
creation operators for the closed strings corresponding to the boundary 
state. 
Let us divide $\left| \Phi \right\rangle$ into the creation 
and annihilation parts as follows:
\begin{eqnarray}
\lefteqn{\left| \Phi \right\rangle
         = \left| \psi  \right\rangle
            + \left| \bar{\psi}  \right\rangle~,}
\nonumber\\
&&\left| \psi \right\rangle
= \int_{0}^{\infty} dl
  \Big[
        \left| n(l) \right\rangle  \phi(l)
          + \bar{\pi}_{0} \left| n(l) \right\rangle  \chi(l)
          +\cdots \Big]~,\nonumber\\
&& \left| \bar{\psi} \right\rangle
= \int^{\infty}_{0} dl
   \Big[
        \left| n(-l) \right\rangle \bar{\phi}(l)
              + \bar{\pi}_{0} \left| n(-l) \right\rangle
                 \bar{\chi}(l)
             + \cdots \Big]~.
 \label{eq:sf2}
\end{eqnarray}

The string field
$\left| \Phi \right\rangle$ satisfies the reality condition:
\begin{equation}
\left\langle \Phi_{\mathrm{hc}} \right|
  = \left\langle \Phi \right|~,
  \label{eq:reality1}
\end{equation}
where
$\left\langle \Phi_{\mathrm{hc}}\right|
\equiv \left(\left|\Phi\right\rangle \right)^{\dagger}$
denotes the hermitian conjugate of
$\left|\Phi\right\rangle$,
and $\left\langle \Phi \right|$ denotes the BPZ conjugate
of $\left|\Phi\right\rangle$
defined in eq.(\ref{eq:BPZ}), respectively.
Since the BPZ conjugation flips the sign of the string
length $\alpha$, the reality condition (\ref{eq:reality1})
implies that
\begin{equation}
\left\langle \psi_{\mathrm{hc}} \right|
  = \left \langle \bar{\psi} \right|~,
\quad
\left\langle \bar{\psi}_{\mathrm{hc}} \right|
  = \left\langle \psi \right|~.
  \label{eq:reality2}
\end{equation}
Combined with the relation
\begin{equation}
\left\langle n(l)_{\mathrm{hc}} \right|
  = \left\langle n(-l) \right|~,
\end{equation}
eq.(\ref{eq:reality2}) leads to
\begin{equation}
\phi^{\dagger}(l) = \bar{\phi}(l)~,
\quad
\chi^{\dagger}(l) = \bar{\chi}(l)~.
\end{equation}

The BRST transformations
$\delta_{\mathrm{B}} \phi(l)$ and
$\delta_{\mathrm{B}} \bar{\phi}(l)$ for 
the component fields $\phi (l)$ and $\bar{\phi}(l)$
can be calculated from eq.(\ref{BRS}). 
Considering the idempotency equations \cite{Kishimoto:2003ru}
 satisfied by the boundary states, 
we expect that the nonlinear terms in the transformation takes a very 
simple form. Indeed we obtain
\begin{eqnarray}
\frac{4C\epsilon^{3}}{g\mathcal{N}}
\delta_{\mathrm{B}} \phi(l)
 &=& \frac{C}{g\mathcal{N}}\epsilon^{2}
     \left(\frac{\partial}{\partial l}
            +\frac{\pi^{2}}{2\epsilon} \right) 
            \left( l\chi(l) \right)
    - \int^{l}_{0} dl_{1}\,
            l_{1}(l-l_{1}) \chi (l_{1}) \phi (l-l_{1})
  \nonumber\\
   && - \int_{0}^{\infty} dl_{1}\,
            l_{1}(l+l_{1}) 
         \left[ \chi(l+l_{1}) \bar{\phi}(l_{1})
                +\bar{\chi}(l_{1}) \phi(l+l_{1}) \right]
      \nonumber\\
  &&  {} + \cdots ~, \nonumber\\
\frac{4C\epsilon^{3}}{g\mathcal{N}}
\delta_{\mathrm{B}} \bar{\phi} (l)
  &=& {}-\frac{C}{g\mathcal{N}}\epsilon^{2}
      \left( \frac{\partial}{\partial l}+\frac{\pi^{2}}{2\epsilon}
      \right) \left( l\bar{\chi}(l) \right)
  +\int^{l}_{0} dl_{1}\,
  l_{1}(l-l_{1}) \bar{\chi}(l_{1}) \bar{\phi}(l-l_{1})
 \nonumber\\
   &&{}  + \int^{\infty}_{0} dl_{1}\,
        l_{1}(l+l_{1})
       \left[ \bar{\chi}(l+l_{1})\phi(l_{1})
             +\chi(l_{1})\bar{\phi}(l+l_{1}) \right]
             \nonumber\\
  && {}+\cdots ~,
  \label{eq:brst}
\end{eqnarray}
where
\begin{equation}
C  = \frac{1}{4\pi^{3}}
    \frac{\left(2\pi^{2}\palpha\right)^{\frac{13}{2}}}
         {\left(4\pi^{2}\palpha\right)^{\frac{p+1}{2}}}
    \sqrt{\frac{V_{\mathrm{N}}}{V_{\mathrm{D}}}}~.
\end{equation}
The derivation of eq.(\ref{eq:brst}) is presented
in Appendix \ref{sec:brst}.
As is intuitively clear, a boundary state split into two makes 
two boundary states and two boundary states joined together makes 
a boundary state. Such contributions are written explicitly 
in eq.(\ref{eq:brst}). However, a boundary state joined to a different 
state makes a state different from the boundary state. 
Such contributions are denoted by `$\cdots$'. 
Notice that each term in `$\cdots$' should
be a product of one annihilation operator 
other than $\phi (l), \chi (l)$ and 
one creation operator other than $\bar{\phi}(l), \bar{\chi}(l)$.


\section{Solitonic Operators}\label{sec:solitonop}
In this section, we will construct solitonic operators made from the 
$\phi$ and $\bar{\phi}$ and study their properties. 

\subsection{Canonical quantization}
Let us canonically quantize the string fields 
defined in eq.(\ref{eq:sf1}) first.
The kinetic term of the action (\ref{eq:actionOSp})
can be written as
\begin{eqnarray}
S_{K}&=&\frac{1}{2} \int dt \int d1 d2
 \left\langle R(1,2)\left| \Phi \right\rangle_{1} \right.
 \left(i\frac{\partial}{\partial t}
     {} - \frac{L_{0}^{(2)}+\tilde{L}_{0}^{(2)}-2}{\alpha_{2}}
     \right)
 \left| \Phi \right\rangle_{2} \nonumber\\
 &=& \int dt \int d1 d2
 \left\langle R(1,2)\left| \bar{\psi} \right\rangle_{1} \right.
 \left(i\frac{\partial}{\partial t}
     {} - \frac{L_{0}^{(2)}+\tilde{L}_{0}^{(2)}-2}{\alpha_{2}}
     \right)
 \left| \psi \right\rangle_{2} \nonumber\\
 &=& \int dt \int  d2
 \,{}_{2}\!\left\langle \bar{\psi}\right|
 \left(i\frac{\partial}{\partial t}
     {} - \frac{L_{0}^{(2)}+\tilde{L}_{0}^{(2)}-2}{\alpha_{2}}
     \right)
 \left| \psi \right\rangle_{2}~.
 \label{eq:osp-kin}
\end{eqnarray} 
In the same way as was performed in the light-cone
string field theory \cite{Kaku:1974zz}\cite{Cremmer:1974ej},
we obtain the canonical commutation relation
\begin{equation}
\left[ \left| \psi \right\rangle_{r},
       {}_{s}\!\left\langle \bar{\psi}\right| \right]
  = I(r,s)
  \qquad \Leftrightarrow \qquad
  \left[ \left|\psi\right\rangle_{r},
       \left|\bar{\psi} \right\rangle_{s} \right]
  =\left| R (r,s) \right\rangle~,
  \label{eq:ccc1}
\end{equation}
where $I(r,s)$ is defined as
\begin{equation}
I(r,s)
= \int du \left\langle R(u,s)\left|R(r,u)\right\rangle\right.~.
\end{equation}
$I(r,s)$ serves as the identity operator.
In fact, the following
relations hold for an arbitrary string field $|\Psi\rangle$,
\begin{equation}
\int ds\, I(r,s) \left| \Psi \right\rangle_{s}
= \left| \Psi \right\rangle _{r}~,
\quad
\int dr \, {}_{r}\! \left\langle \Psi \right|I(r,s)
  ={}_{s}\! \left\langle \Psi \right|~.
\end{equation}
Multiplying the second equation in eq.(\ref{eq:ccc1})
by $\int dr\, {}_{r}\! \langle n(-l_{r}) |
     \int ds \, {}_{s}\! \langle n(l_{s}) |$
{}from the left, we have
\begin{equation}
\left[ \phi(l_{r}) , \bar{\phi}(l_{s}) \right]
 = \frac{1}{\mathcal{N}^{2} l_{r}} \delta (l_{r}-l_{s})~.
 \label{eq:ccc2}
\end{equation}
One can also derive this commutation relation
directly from the action (\ref{eq:osp-kin})
expressed in terms of the component fields:
\begin{eqnarray}
S_{K}= \mathcal{N}^2 \int dt \int^{\infty}_{0} dl\,
     l\,\bar{\phi}(l) i \frac{\partial \phi(l)}{\partial t}
        {}+ \cdots~.
\end{eqnarray}

The vacuum state
$|0\rangle\!\rangle$ in the
second quantization is defined by
\begin{equation}
|\psi \rangle |0\rangle\!\rangle =0~,
\quad
\langle\!\langle 0 | \langle \bar{\psi} | =0~.
\end{equation}
This yields
\begin{equation}
\phi (l) |0\rangle \! \rangle
    = \chi(l) |0\rangle\!\rangle =0~,
\quad \langle \! \langle 0| \bar{\phi} (l)
   = \langle \! \langle 0| \bar{\chi} (l) =0,
\qquad \mbox{for $l>0$}~.
\end{equation}
We take the normalization of the vacuum state
$|0\rangle\!\rangle$ as
$\langle \! \langle 0 | 0 \rangle \! \rangle =1$.

\subsection{Solitonic states}
The right hand sides of eq.(\ref{eq:brst}) look quite like those of 
eq.(\ref{noncriticalcommutation}). Indeed if we replace 
\begin{eqnarray}
&\displaystyle
 \left[ \, \frac{1}{g_s^2}\int^{\infty}_{0} dl^\prime
            l^\prime \epsilon (l^\prime )
            T(l^\prime )\, , \, \ \cdot \  \,\right]
  \ \rightarrow
  \  \frac{4C\epsilon^{3}}{g\mathcal{N}}\delta_{\mathrm{B}}
    \left( \ \cdot \ \right)~,&\nonumber\\
&\displaystyle
  \frac{\sqrt{2}}{g_s\mathcal{N}l}\psi (l)
   \rightarrow \phi (l)~,
  \quad
\frac{g_s}{\sqrt{2}\mathcal{N}}\psi^\dagger (l)
   \rightarrow \bar{\phi}(l)~,
  \quad 
\epsilon (l)
   \rightarrow \bar{\chi}(l)~,&
\end{eqnarray}
in eq.(\ref{noncriticalcommutation}),
we get exactly the nonlinear terms involving $\bar{\chi}(l)$ on the 
right hand sides of eq.(\ref{eq:brst}). 
Moreover the commutation relation (\ref{noncriticalcanonical})
becomes eq.(\ref{eq:ccc2}) by such replacements. 
Therefore we expect that operators in the following form can be
used to construct BRST invariant operators:
\begin{equation}
\exp\left[
   \pm \sqrt{2}\mathcal{N} \int^{\infty}_{0} dl\,
     \,e^{-\zeta l} \bar{\phi}(l)\right]
    \,\exp \left[
      \mp \sqrt{2}\mathcal{N} \int^{\infty}_{0} dl'
    \, e^{\zeta l'} \phi (l')\right].
\end{equation}
Taking the linear terms on the right hand sides of eq.(\ref{eq:brst}) 
into account, we define 
\begin{eqnarray}
\mathcal{V}(\zeta)
= \lambda \exp\left[
   \pm \sqrt{2}\mathcal{N} \int^{\infty}_{0} dl\,
     \,e^{-\zeta l} \bar{\phi}(l)\right]
    \,\exp \left[
      \mp \sqrt{2}\mathcal{N} \int^{\infty}_{0} dl'
    \, e^{\zeta l'} \phi (l')\right]
    e^{\pm \frac{C\epsilon^{2}}{\sqrt{2}g}
        \left(\zeta+\frac{\pi^{2}}{2\epsilon}\right)^{2}}~,
\label{eq:v-cal}
\end{eqnarray}
where $\lambda$ is a constant. 
Actually we cannot make BRST invariant
operators from $\mathcal{V}(\zeta )$. 
Rather we will show that
we can construct BRST invariant states
by acting
$\int d\zeta \mathcal{V}(\zeta)$ on the 
vacuum $|0\rangle\!\rangle$. 

As a warm-up, let us show that 
\begin{eqnarray}
\left| D \right\rangle \! \rangle
&\equiv&
\int d\zeta\, \mathcal{V}(\zeta) |0\rangle\!\rangle
\nonumber
\\
&=&
\lambda
\int d\zeta \exp \left[
                \pm  \sqrt{2}\mathcal{N} \int^{\infty}_{0} dl\,
                e^{-\zeta l}\bar{\phi}(l) 
                \pm \frac{C\epsilon^{2}}{\sqrt{2}g}
                    \left(\zeta + \frac{\pi^{2}}{2\epsilon} 
                    \right)^{2} \right]
     |0 \rangle \! \rangle
    \label{eq:d-branestate}
\end{eqnarray}
is BRST invariant:
\begin{equation}
\delta_{\mathrm{B}} |D\rangle \! \rangle =0~.
\label{eq:brsinv-dstate-0}
\end{equation}
The BRST transformation can be calculated by using eq.(\ref{eq:brst}) 
as
\begin{eqnarray}
\lefteqn{
  \frac{4C\epsilon^{3}}{g\mathcal{N}}
   \,\delta_{\mathrm{B}}\left(
  e^{\pm \sqrt{2}\mathcal{N} \int^{\infty}_{0} dl\, e^{-\zeta l}
            \bar{\phi}(l) } \right)
  |0\rangle \!\rangle
}  \nonumber\\
&&\ = \left[ \mp \frac{C\epsilon^{2}\sqrt{2}}{g}
      \int^{\infty}_{0} dl\, e^{-\zeta l}
      \left(\frac{\partial}{\partial l} + \frac{\pi^{2}}{2\epsilon}
      \right) \left( l \bar{\chi}(l) \right) 
    \right.\nonumber\\
 && \hspace{2.5em} 
    {}\pm \sqrt{2} \mathcal{N} \int^{\infty}_{0} dl\,
     e^{-\zeta l} \int^{l}_{0} dl_{1}\, l_{1}(l-l_{1})
        \bar{\chi}(l_{1}) \bar{\phi}(l-l_{1}) 
     \nonumber\\
&& \hspace{2.5em}
    \left.
    {}+
      \int^{\infty}_{0} dl \int^{\infty}_{0} dl_{1}\,
        e^{-\zeta (l+l_{1})}
         (l+l_{1}) \bar{\chi}(l+l_{1}) \right]    
    \,e^{\pm \sqrt{2} \mathcal{N} \int^{\infty}_{0} dl\, e^{-\zeta l}
            \bar{\phi}(l) } 
  \; |0\rangle \!\rangle~.
    \label{eq:brstexp1}
\end{eqnarray}
Here we have used
$\phi(l)|0\rangle\!\rangle=\chi(l)|0\rangle\!\rangle=0$ and 
the fact that `$\cdots$' in eq.(\ref{eq:brst}) does not contribute 
because it includes annihilation operators other than $\phi$. 
It is useful to
introduce the Laplace transforms $\tilde{\bar{\phi}}(\zeta)$
and $\tilde{\bar{\chi}}(\zeta)$ of
$\bar{\phi}(l)$ and $\bar{\chi}(l)$ defined as
\begin{equation}
\tilde{\bar{\phi}}(\zeta) = \int^{\infty}_{0} dl \,
    e^{-\zeta l}\bar{\phi}(l)~,
\qquad
\tilde{\bar{\chi}}(\zeta) = \int^{\infty}_{0} dl \,
    e^{-\zeta l}\bar{\chi} (l)~.
\label{eq:laplace}
\end{equation}
The following identities hold,
\begin{eqnarray}
\int^{\infty}_{0} dl\, e^{-\zeta l}
  \left( \frac{\partial}{\partial l}+\frac{\pi^{2}}{2\epsilon}
  \right) \left( l \bar{\chi} (l) \right)
 &=& -\left( \zeta + \frac{\pi^{2}}{2\epsilon} \right)
    \frac{\partial}{\partial \zeta} \tilde{\bar{\chi}} (\zeta)~,
    \nonumber\\
\int^{\infty}_{0} dl\, e^{-\zeta l}
   \int^{l}_{0} dl_{1}\, l_{1}(l-l_{1}) \bar{\chi}(l_{1})
   \bar{\phi}(l-l_{1})
   &=&\frac{\partial \tilde{\bar{\chi}}(\zeta)}{\partial \zeta}
    \frac{\partial \tilde{\bar{\phi}}(\zeta)}{\partial \zeta}~,
   \nonumber\\
\int^{\infty}_{0} dl\, e^{-\zeta l}
   \int^{\infty}_{0} dl_{1}\, e^{-\zeta l_{1}}
    (l+l_{1}) \bar{\chi}(l+l_{1})
   &=& \frac{\partial^{2}}{\partial \zeta^{2}}
     \tilde{\bar{\chi}}(\zeta)~.
    \label{eq:FS}
\end{eqnarray}
Combining these relations with eq.(\ref{eq:brstexp1}),
we obtain
\begin{eqnarray}
\lefteqn{\frac{4C\epsilon^{3}}{g\mathcal{N}} \,
  \delta_{B}\left(
  \exp\left[\pm \sqrt{2} \mathcal{N}
            \int^{\infty}_{0}dl\, e^{-\zeta l}\bar{\phi}(l)
      \pm \frac{C\epsilon^{2}}{\sqrt{2}g}
        \left( \zeta + \frac{\pi^{2}}{2\epsilon} \right)^{2}
      \right]
  \right) |0\rangle\!\rangle 
 }\nonumber\\
&&  = \frac{\partial}{\partial \zeta}
       \left( \frac{\partial \tilde{\bar{\chi}}(\zeta)}{\partial \zeta}
       \exp\left[\pm \sqrt{2} \mathcal{N} 
       \int^{\infty}_{0}dl\, e^{-\zeta l}\bar{\phi}(l)
      \pm \frac{C\epsilon^{2}}{\sqrt{2}g}
        \left( \zeta + \frac{\pi^{2}}{2\epsilon} \right)^{2}\right]
        \right)\,|0\rangle\!\rangle~.
     \label{eq:brstexp3}
\end{eqnarray}
Taking eq.(\ref{eq:d-branestate}) into account,
we find that this equation implies
the BRST invariance (\ref{eq:brsinv-dstate-0}) of
the state $|D\rangle\!\rangle$.

$|D\rangle\!\rangle$ can be considered as a state in which D-branes are 
excited. Actually as we will see in the next subsection, two D-branes are 
there. In order to have more D-branes, we just have to operate 
$\int d\zeta \mathcal{V}(\zeta)$ 
successively on the vacuum $|0\rangle\!\rangle$,
namely we construct states
\begin{equation}
\left(\int d\zeta \mathcal{V}(\zeta)\right)^n|0\rangle\!\rangle ~,
\end{equation}
for $n>0$. They are BRST invariant,
because
$\int d\zeta \mathcal{V}(\zeta)$
is BRST invariant 
modulo terms which annihilate the states 
$(\int d\zeta \mathcal{V}(\zeta))^n|0\rangle\!\rangle$,
$n\geq 0$.
This can be seen as follows.
Under the BRST transformation~(\ref{eq:brst}),
$\mathcal{V}(\zeta)$ transforms as
\begin{equation}
\frac{4C\epsilon^{3}}{g\mathcal{N}}\,
\delta_{\mathrm{B}} \mathcal{V}(\zeta)
= \frac{\partial}{\partial \zeta}
  \left[ \left( 
   \frac{\partial \tilde{\bar{\chi}}(\zeta)}{\partial \zeta}
   - \frac{\partial \tilde{\chi} (\zeta)}{\partial \zeta}\right)
    \mathcal{V}(\zeta) \right]
  +\left( \pm \frac{2\sqrt{2}C\epsilon^{2}}{g}
           \zeta \frac{\partial \tilde{\chi}(\zeta)}{\partial \zeta}
         +\cdots \right) \mathcal{V}(\zeta)~,
       \label{eq:brst-v}
\end{equation}
where `$\cdots$' denotes the terms which includes annihilation 
operators other than $\phi(l)$,
and $\tilde{\phi}(\zeta)$ and $\tilde{\chi}(\zeta)$
are the Laplace transforms of $\phi(l)$ and $\chi(l)$ defined as
\begin{equation}
\tilde{\phi}(\zeta) = \int^{\infty}_{0} dl\,
   e^{\zeta l} \phi(l)~,
   \quad
\tilde{\chi}(\zeta) = \int^{\infty}_{0} dl\, e^{\zeta l}\chi(l)~.
\end{equation}
We can prove
eq.(\ref{eq:brst-v}) with the help of the relation
\begin{eqnarray}
\lefteqn{
 \int^{\infty}_{0} dl\, e^{-\zeta l}
 \int^{\infty}_{0} dl_{1}
 \, l_{1}(l+l_{1}) \chi(l_{1})\bar{\phi}(l+l_{1})
}\nonumber\\
&&  = -\frac{\partial \tilde{\chi}(\zeta)}{\partial \zeta}
     \frac{\partial \tilde{\bar{\phi}}(\zeta)}{\partial \zeta}
    -\int^{\infty}_{0} dl \int^{\infty}_{0}dl_{1}
      \, e^{\zeta l} l_{1}(l+l_{1}) \chi(l+l_{1})\bar{\phi}(l_{1})~.
\end{eqnarray}
The terms in the parenthesis $(\quad)$ in the second term
on the right hand side of eq.(\ref{eq:brst-v})
commute with $\mathcal{V}(\zeta)$ and annihilate
$|0\rangle\!\rangle$.
Thus we find that the states of the form 
$(\int d\zeta \mathcal{V}(\zeta))^n|0\rangle\!\rangle$
are BRST invariant. 

\subsection{Vacuum amplitude}
In order to compare our description of D-branes
constructed above
with the usual one, let us 
compute the vacuum amplitude in the presence of D-branes in our 
formalism. 
Since $|D\rangle\!\rangle$ is considered as a state with some 
D-branes excited, the vacuum amplitude in the presence of 
D-branes can be given as 
\begin{equation}
\lim_{T\rightarrow\infty}
\langle\!\langle D| e^{-i T\hat{H}} |D\rangle\!\rangle~,
\end{equation}
where $\hat{H}$ is the second-quantized Hamiltonian. 
Notice that the time variable in the $OSp$ invariant string field 
theory is not the physical time but with ``topological" nature 
because the Hamiltonian $\hat{H}$ is BRST exact.
This can be seen from the expression (\ref{eq:BRSTexact}).
Therefore $\langle\!\langle D| e^{-i T\hat{H}} |D\rangle\!\rangle$ can 
be transformed into a form which is independent of the value of $T$. 
By comparing $\langle\!\langle D| e^{-i T\hat{H}} |D\rangle\!\rangle$
with the usual vacuum
amplitude,
we can see how many D-branes are there
in the state $|D\rangle\!\rangle$.

In order to do so, let us first perform the integration over 
$\zeta$ in  eq.(\ref{eq:d-branestate}). 
Perturbatively the factor 
\begin{equation}
\exp \left[\pm \frac{C\epsilon^{2}}{\sqrt{2}g}
    \left(\zeta + \frac{\pi^{2}}{2\epsilon} \right)^{2}\right]
  \label{eq:gaussian}
\end{equation}
in the integrand in eq.(\ref{eq:d-branestate}) is the most dominant.
Therefore,
through the saddle point approximation we obtain
\begin{equation}
|D\rangle \! \rangle
  \simeq \lambda'
  \exp\left[ \pm  \sqrt{2}\mathcal{N} 
       \int^{\infty}_{0} dl\,
        e^{\frac{\pi^{2}}{2\epsilon}l}
       \bar{\phi} (l) \right] |0\rangle \! \rangle~,
    \label{eq:saddle}
\end{equation}
where
$\lambda' 
 = \sqrt{\mp\frac{\sqrt{2}\pi g}{ C\epsilon^{2}}}
 \lambda$.
One can find that
\begin{eqnarray}
\bar{\phi}(l) \, e^{\frac{\pi^{2}}{2\epsilon}l }
 &=& -\frac{1}{\mathcal{N}^{2} l}
      \int dr\, e^{\frac{\pi^{2}}{2\epsilon}l}\,
          {}_{r}\! \langle n(l) | \bar{\psi} \rangle_{r}
   \nonumber\\
 &=& -\frac{1}{\mathcal{N}^{2}}
      \int dr\, \frac{1}{l}\,
    {}^{\epsilon}_{r}\! \langle B_{0} (l) |
                          \bar{\psi} \rangle_{r}
= \frac{1}{\mathcal{N}^{2}}\int dr\, \frac{1}{l}
  {}_{r}\!\langle \bar{\psi} | B_{0}(l)
          \rangle^{\epsilon}_{r}~.
\end{eqnarray}
Plugging this relation into eq.(\ref{eq:saddle}),
we obtain the expression of the state $| D \rangle\!\rangle$
in terms of the string state $\left| \psi \right\rangle$ and
$|B_{0}(l)\rangle^{\epsilon}$ as follows:
\begin{equation}
|D\rangle \! \rangle = \lambda'
 \exp \left[ \pm 
     \int dr \int^{\infty}_{0} \frac{dl}{l}
     \ \frac{\sqrt{2}}{\mathcal{N}} \;
     {}_{r}\! \langle \bar{\psi} |
                B_{0}(l) \rangle^{\epsilon}_{r}
     \right] |0\rangle \! \rangle~.
   \label{eq:d-branestate2}
\end{equation}
Note that the divergent factor $e^{\frac{\pi^{2}}{2\epsilon} l}$
$(l>0)$
is miraculously canceled by the regularization factor
$e^{-\frac{\pi^{2}}{2\epsilon}|l|}$ in $|n(l)\rangle$ 
and we can express $|D\rangle \! \rangle $ in terms of 
$|B_0(l)\rangle$.

Now let us evaluate 
$\langle\!\langle D| e^{-i T\hat{H}} |D\rangle\!\rangle$. 
Perturbatively the lowest order contribution can be obtained 
by replacing the Hamiltonian $\hat{H}$ with its free part
$\hat{H}_{0}$. 
Substituting
\begin{eqnarray}
\langle\!\langle D|  &=&
    \lambda^{\prime \dagger}
    \langle\!\langle 0 |
       \exp \left[ \pm \int dr \int^{\infty}_{0} \frac{dl}{l}
          \frac{\sqrt{2}}{\mathcal{N}}
          \, {}_{r}^{\epsilon} \langle
            B_{0} (-l) | \psi \rangle_{r} \right]~,
\nonumber\\
\hat{H}_{0}
  &=& \int dr\, {}_{r}\!\langle \bar{\psi}|
    \, \frac{L^{(r)}_{0}+\tilde{L}^{(r)}_{0}-2}{\alpha_{r}}
    \,  \left| \psi \right\rangle_{r}~,
\end{eqnarray}
into the expression and 
using the commutation relation (\ref{eq:ccc1}), we find that
\begin{eqnarray}
\lefteqn{
 \langle\!\langle D | e^{-i T\hat{H}_{0}} | D\rangle\!\rangle}
  \nonumber\\
&&= \left| \lambda' \right|^{2}
   \langle \! \langle 0 |
   \exp \left[  \frac{2}{\mathcal{N}^{2}}
  \int dr \int^{\infty}_{0} \frac{dl}{l}\;
    {}^{\epsilon}_{r}\! \langle B_{0}(-l)| \psi \rangle_{r}
    \ e^{-i T\hat{H}_{0}}  \  
    \int dr' \int^{\infty}_{0} \frac{dl'}{l'} \;
    {}_{r'}\! \langle \bar{\psi} |
       B_{0}(l') \rangle^{\epsilon}_{r'}
     \right]  |0\rangle\!\rangle   \nonumber\\
&&=  \left| \lambda' \right|^{2}
   \exp \left[
       \frac{2}{\mathcal{N}^{2}} \int^{\infty}_{0}\frac{dl}{l}
                \int^{\infty}_{0} \frac{dl'}{l'}
      \int dr \;
        {}^{\epsilon}_{r}\! \langle B_{0}(-l')|\,
         e^{-iT\frac{L^{(r)}_{0}+\tilde{L}^{(r)}_{0}-2}{\alpha_{r}}}
         \, |B_{0}(l) \rangle^{\epsilon}_{r}
       \right]~.
      \label{eq:dd1}
\end{eqnarray}
Here we have used the relation
\begin{equation}
\langle \! \langle 0 |
 \; {}^{\epsilon}_{r}\!\left\langle B_{0} (-l) \left.\right|
   \psi \right\rangle_{r} \, \hat{H}_{0}
 = \langle \! \langle 0| \; {}^{\epsilon}_{r}\!
   \left\langle B_{0} (-l) \right|
    \frac{L^{(r)}_{0}+\tilde{L}_{0}^{(r)}-2}{\alpha_{r}}
    \left| \psi \right\rangle_{r}~,
\end{equation}
which follows from eq.(\ref{eq:ccc1}).

The quantity in the exponent of eq.(\ref{eq:dd1}) should 
be compared with the cylinder amplitude in the usual formulation. 
After the integrations over $\alpha$, $\pi_0$, $\bar{\pi}_0$ and 
$l^\prime$, we obtain the integration measure for $l$ as 
\begin{equation}
\int_0^\infty dl\frac{T}{l^2}
 =\int_0^\infty d \left( \frac{T}{l} \right)~.
\end{equation}
Since $T$ is fixed, this is exactly the integration over the parameter 
in front of $L_0+\tilde{L}_0-2$. 
Therefore the integration over $l$ is transformed into the one for the 
moduli parameter of the cylinder and the result is in a form which 
is independent of the value of $T$. 
The overlap between the boundary states
in the exponent on the right hand side of eq.(\ref{eq:dd1})
can readily be obtained from eq.(\ref{eq:overlap2})
by replacing $\epsilon$ in eq.(\ref{eq:overlap2}) by $i T/2$.
Introducing 
$\tau^\prime \equiv - \pi \frac{l}{T}$,
we obtain
\begin{eqnarray}
\lefteqn{\langle\!\langle D | e^{-i T\hat{H}_{0}} 
     | D \rangle \! \rangle
}\nonumber\\
&&= \left| \lambda' \right|^{2}
  \exp \left[
  4 \int^{\infty}_{0} \frac{d\tau^\prime}{2\tau^\prime} \eta(-\tau^\prime )^{-24}
    \prod_{\mu \in \mathrm{N}}
    \left( \sum_{m \in \mathbb{Z}} 
         e^{-i\frac{2\pi\palpha}{(R^{\mu})^{2}} \tau^\prime m^{2}} \right)
     \prod_{i\in\mathrm{D}}\left( \sum_{n \in \mathbb{Z}}
         e^{-i\frac{2\pi(R^{i})^{2}}{\palpha} \tau^\prime n^{2}}
     \right)\right]~.
   \label{eq:dd2}
\end{eqnarray}
Notice that the arbitrary normalization constant $\mathcal{N}$ 
does not appear in this final answer. 
We can see that the exponent of eq.(\ref{eq:dd2})
reproduces four times the annulus amplitude for one D-brane.
This result implies that $|D\rangle\!\rangle$ 
is the state in which two D-branes or ghost D-branes are excited. 
Since $\bar{\phi}$ generates boundaries on the worldsheet, 
depending on which sign in eq.(\ref{eq:v-cal}) is chosen, 
D-branes or ghost D-branes are excited. 
Similar calculations yield that $\int d\zeta \mathcal{V}(\zeta)$ 
creates two D-branes or ghost D-branes. 

In this subsection, we have shown that the cylinder amplitudes 
for D-branes are reproduced in our formulation. What is remarkable 
is that the integration measure for the moduli of the cylinder 
appear from the integration over the length of the string. 
It would be an intriguing problem to check if the integration measures 
for the higher genus graphs appear in a similar way.

\section{Discussions}\label{sec:discussions}

In this paper, we have constructed solitonic operators which create 
D-branes. Although we started from
the nonnormalizable state (\ref{eq:B0}),
the  divergent factors cancel with each other
and the cylinder amplitude 
is reproduced. The cancellation occurred because of the factor 
$\exp \left[\pm \frac{C\epsilon^{2}}{\sqrt{2}g}
    \left(\zeta + \frac{\pi^{2}}{2\epsilon} \right)^{2}\right]$ 
in $\mathcal{V}(\zeta )$. It originates from the term 
$-i\pi_0\partial_\alpha$ in the BRST charge~(\ref{eq:brst-charge}). 
This term is peculiar to the $OSp$ invariant theory and it seems 
that our construction works only for this theory. 
The exponent of the  exponential factor we are discussing
may be interpreted as the potential for the open string 
tachyon. Indeed $\zeta$ appears in the form of
$\exp (-\zeta l)$ in front of 
$\phi(l)$ and can be considered as a constant tachyon background. 
It will be useful to interpret various quantities in our construction 
in terms of open string language using for example the methods in 
Ref.\cite{Murakami:2002yd}. 


In our construction,
we do not describe D-branes as solutions of equations of 
motion. Rather we construct the solitonic operator 
$\int d\zeta \mathcal{V}(\zeta )$,
 where the form of $\mathcal{V}(\zeta )$ 
looks quite like the bosonization formula. 
Another way to look at our results is as follows. 
We may write
\begin{equation}
\int^{\infty}_{0} \frac{dl}{l}\;
   {}^{\epsilon}_{r}\!\left\langle B_{0}(l)\left|
     \bar{\psi}\right\rangle_{r}\right.
 =\int^{\infty}_{-\infty} \frac{dl}{l}\;
   {}^{\epsilon}_{r}\!\left\langle B_{0}(l)\left|
     \bar{\psi}\right\rangle_{r}\right.
 ={}^{\epsilon}_{r} \left\langle B \left| \bar{\psi}
                    \right\rangle_{r}
   \right.~,
\end{equation}
by introducing
the state $\left|B\right\rangle^{\epsilon}$
defined as
\begin{equation}
\left| B \right\rangle^{\epsilon}
  = \left| B_{0} \right\rangle^{\epsilon}\frac{1}{\alpha}~.
  \label{eq:invbs}
\end{equation}
This yields
\begin{equation}
|D\rangle\!\rangle = \lambda'\exp\left[
   \mp \frac{\sqrt{2}}{\mathcal{N}} \int dr\,
    {}^{\epsilon}_{r}\!\left\langle B \left| \bar{\psi} 
    \right\rangle_{r} \right.
    \right]\;
    |0\rangle\!\rangle~.
    \label{eq:d-b-bpsi}
\end{equation}
The state $\left| B \right\rangle^{\epsilon}$
is regarded as a regularized version of the state
$
\left| B \right\rangle \equiv \left| B_{0} \right\rangle
  \frac{1}{\alpha}$.
We find that the states $\left| B \right \rangle$
and $\left| B \right \rangle^{\epsilon}$
are annihilated by the BRST charge $Q_{\mathrm{B}}$:
\begin{equation}
Q_{\mathrm{B}} \left| B \right\rangle
=Q_{\mathrm{B}} \left| B \right\rangle^{\epsilon} =0~.
\label{eq:qbb}
\end{equation}
We can extend eq.(\ref{eq:d-b-bpsi})
by including the dependence on the annihilation
modes $|\psi\rangle$ into
\begin{equation}
|D\rangle \! \rangle =
 \lambda'
 \; : \exp\left[
  \mp  \frac{\sqrt{2}}{\mathcal{N}}\int dr\;
     {}^{\epsilon}_{r}\!\left\langle \left. B\right|
     \Phi \right\rangle_{r} \right]:
    |0\rangle\! \rangle~,
  \label{eq:d-branestate3}
\end{equation}
where $:\ :$ means the normal ordering in which
the annihilation mode $|\psi\rangle$ should be
moved to the right of the creation mode
$|\bar{\psi}\rangle$.
Taking account of the relation (\ref{eq:qbb}),
we notice that the exponent of eq.(\ref{eq:d-branestate3})
takes a form quite similar to
the interaction term of a closed string
with a D-brane introduced by Ref.\cite{Hashimoto:1997vz}
into the action of
the HIKKO closed string field theory~\cite{Hata:1986kj}.
Therefore we can say that 
D-branes are introduced as a source of closed strings.

As we mentioned, we should introduce a cut-off $\delta$ so that 
$|\alpha |,|l| >\delta$. 
Since $l$ works as a moduli, this cut-off will 
remove the singularity occurring in the limit
where a part of the worldsheet becomes a very long cylinder.
Therefore every equation in this paper
should be understood with such a cut-off being
introduced.
In the first relation in eq.(\ref{eq:FS}),
we have ignored
the contribution
from the boundary term
$ e^{-\zeta l}l\bar{\chi}(l) \Big|_{l=0}$.
Due to the cut-off, we should take care of
the contribution of the boundary term near $l=0$,
which violates the BRST invariance of the state
$|D\rangle\!\rangle$.
We should therefore employ
the Fischler-Susskind mechanism~\cite{Fischler:1986ci},
in order to remedy the violation of the BRST
invariance.
This modifies the equations of motion
for light states of closed string. 
This is also consistent with the picture of
the D-brane as a source of closed strings.
Our treatment in section \ref{sec:solitonop}
is perturbative and ignores such higher order effects.

There are several problems that remain to be studied. 
One immediate question is why 
our solitonic operators create two D-branes or ghost D-branes. We 
are not sure if there exist operators which create one D-brane 
or ghost D-brane. 
It will be an intriguing problem to look for such operators. 
In this paper, we have not fixed the sign
in the exponents of eqs.(\ref{eq:v-cal}) and
(\ref{eq:d-branestate}).
{}From the calculation (\ref{eq:dd2}),
one can find that if the state $|D\rangle\!\rangle$
with one sign is identified with ordinary D-branes,
the state with the other sign should be
identified with ghost D-branes
\cite{Takayanagi}.
To determine
which sign corresponds to which will be
another interesting problem.
Of course, our results should be generalized to the superstring case. 
In order to do so, we should first construct the $OSp$ invariant 
theory for superstrings.

\section*{Acknowledgements}
We would like to thank K.~Araki, N.~Hatano, S.~Katagiri
and T.~Saitou for discussions.
This work is supported in part
by Grants-in-Aid for Scientific Research 13135224.

\appendix

\section{Derivation of eq.(\ref{eq:brst})}\label{sec:brst}

In this appendix, we provide some details of the calculation to 
get eq.(\ref{eq:brst}) from
the BRST transformation (\ref{BRS})
for the string field $| \Phi\rangle$ given in eq.(\ref{eq:sf1}).
For later convenience, we introduce the notation
\begin{equation}
\bphi (l) = \left\{ \begin{array}{ll}
               \phi (l) & \mbox{for $l>0$}\\
               \bar{\phi}(-l) & \mbox{for $l<0$}
                    \end{array} \right.~,
\quad
\bchi(l) = \left\{ \begin{array}{ll}
                \chi(l) & \mbox{for $l>0$}\\
                \bar{\chi}(-l) & \mbox{for $l<0$}
                   \end{array} \right.~.
\end{equation}

In order to get $\delta_{\mathrm{B}} \phi$ and 
$\delta_{\mathrm{B}} \bar{\phi}$, we consider the inner 
product of $\left\langle n(-l) \right|$ and eq.(\ref{BRS}). 
As for the left hand side of the BRST transformation,
one can readily find that
\begin{equation}
n(-l) \cdot \delta_{\mathrm{B}}\Phi
\equiv
\int dr \, {}_{r}\!
  \left\langle n(-l) \left| \delta_{\mathrm{B}}\Phi
  \right\rangle_{r} \right.
  = \mathcal{N}^{2} l \, \delta_{\mathrm{B}} \bphi (l)
  = \left\{
    \begin{array}{ll}
      \mathcal{N}^{2} l \delta_{\mathrm{B}}\phi (l) &
           \mbox{for $l>0$} \\
      \mathcal{N}^{2} l \delta_{\mathrm{B}} \bar{\phi} (-l) &
           \mbox{for $ l<0$}
     \end{array}
    \right.~.
  \label{eq:brs-lhs}
\end{equation}

Let us turn to the right hand side of the BRST transformation.
First, we consider the linear term $Q_{\mathrm{B}}|\Phi\rangle$.
Using eq.(\ref{eq:reflection2}), we find that
\begin{eqnarray}
n(-l) \cdot Q_{\mathrm{B}}\Phi
 &\equiv&
 \int d1 d2 \langle R(1,2) | n(-l) \rangle_{1}
             Q_{\mathrm{B}}^{(2)}|\Phi\rangle_{2}
 \nonumber\\
&=& -\int d1 d2 \langle R(1,2) |Q_{\mathrm{B}}^{(1)}
      | n(-l) \rangle_{1}  |\Phi\rangle_{2}~.
   \label{eq:nqb-1}
\end{eqnarray}
The state $|n(-l)\rangle$ is expressed as
\begin{equation}
\left| n(-l) \right\rangle
 =\left| B \right\rangle^{\epsilon}
  e^{-\frac{\pi^{2}}{2\epsilon} |l| }
  \alpha \delta (\alpha + l)~,
 \label{eq:n-B}
\end{equation}
where $\left| B \right\rangle^{\epsilon}$ is
the state introduced
in eq.(\ref{eq:invbs}).
Combined with eq.(\ref{eq:qbb}),
this leads to
\begin{eqnarray}
Q_{\mathrm{B}} |n(-l) \rangle
&=& Q_{\mathrm{B}}
    \left(
    \left| B \right\rangle^{\epsilon}
    e^{-\frac{\pi^{2}}{2\epsilon} |l|}
    \alpha \delta(\alpha+l)  \right)
= \left| B \right\rangle^{\epsilon}
    e^{-\frac{\pi^{2}}{2\epsilon} |l|}
   Q_{\mathrm{B}}\left(
       \alpha \delta(\alpha+l) \right)\nonumber\\
&=& \left| B \right\rangle^{\epsilon}
     e^{-\frac{\pi^{2}}{2\epsilon} |l|}
    (-i\pi_{0})\frac{\partial}{\partial \alpha}
    \left( \alpha     \delta(\alpha+l) \right)
=\left| B \right\rangle^{\epsilon}
     e^{-\frac{\pi^{2}}{2\epsilon} |l|}
    i\pi_{0} l \frac{\partial}{\partial l}
     \delta(\alpha+l)~.
\end{eqnarray}
Substituting this equation into eq.(\ref{eq:nqb-1}),
we have
\begin{eqnarray}
n(-l) \cdot Q_{B} \Phi
 &=& - i e^{-\frac{\pi^{2}}{2\epsilon} |l|}
     l \int d1 d2 \, \langle R(1,2) | \pi^{(1)}_{0}
      \frac{\partial}{\partial l}\delta(\alpha_{1}+l)
      \left|B\right\rangle^{\epsilon}_{1} 
      \left|\Phi \right\rangle_{2}
   \nonumber\\
 &=& i e^{-\frac{\pi^{2}}{2\epsilon} |l|}
     l \int d1 d2 \, \langle R(1,2) | 
      \frac{\partial}{\partial l}\delta(\alpha_{1}+l)
      \left|B\right\rangle^{\epsilon}_{1} 
      \pi^{(2)}_{0}\left|\Phi \right\rangle_{2}
     \nonumber\\
  &=& i e^{-\frac{\pi^{2}}{2\epsilon} |l|}
     l \int d1 d2 \,
     \frac{\partial}{\partial l}\left(
        \left\langle R(1,2) \right| \delta(\alpha_{1}+l)
          \left|B\right\rangle^{\epsilon}_{1} \pi^{(2)}_{0}
          \left| \Phi \right\rangle_{2}  \right)
      \nonumber\\
  &=& i e^{-\frac{\pi^{2}}{2\epsilon} |l|}
     l \int d1 d2 \,
     \frac{\partial}{\partial l}\left(
        \left\langle R(1,2) \right|
         \frac{-1}{l}
          e^{\frac{\pi^{2}}{2\epsilon} |l|}
         \left| n(-l) \right\rangle_{1}
         \pi^{(2)}_{0} \left| \Phi \right\rangle_{2} 
     \right)  \nonumber\\
  &=& -i e^{-\frac{\pi^{2}}{2\epsilon} |l|}
      l \frac{\partial}{\partial l}\left(
        \frac{1}{l}
       e^{\frac{\pi^{2}}{2\epsilon} |l|}
     \int d2 \,
       {}_{2}\!\left\langle n(-l) \right|
         \pi^{(2)}_{0} \left| \Phi \right\rangle_{2}
     \right)~.
  \label{eq:nqb-2}
\end{eqnarray}
On the rightest hand side in the above equation,
only the $\bar{\pi}_{0} \left|n(l')\right\rangle  \bchi(l')$ 
component of $\left| \Phi \right\rangle$ provides 
a nonvanishing 
contribution because of the ghost zero-mode
saturation, i.e.
\begin{equation}
\int d2\, {}_{2}\! \left\langle n(-l)\right|
    \pi^{(2)}_{0} \left|\Phi\right\rangle_{2}
= \int^{\infty}_{-\infty} dl'
  \int d2 {}_{2}\! \left\langle n(-l) \right|
       \pi^{(2)}_{0} \bar{\pi}^{(2)}_{0}
       \left| n(l') \right\rangle_{2}  \bchi (l')~.
     \label{eq:nbpiphi}
\end{equation}
It is easy to show 
\begin{equation}
\int d2\,{}_{2}\! \left\langle n(-l) \right|
       \pi^{(2)}_{0} \bar{\pi}^{(2)}_{0}
       \left| n(l') \right\rangle_{2}
= \frac{i|l|}{4\epsilon}
   \int d2\,{}_{2}\! \left\langle n(-l) \right|
       \left. n(l') \right\rangle_{2}
= i\frac{\mathcal{N}^{2}}{4\epsilon}
   \, |l| \,l' \,\delta(l'-l)~.
\label{eq:n-pipi-n}
\end{equation}
Plugging eqs.(\ref{eq:n-pipi-n})  and (\ref{eq:nbpiphi})
into eq.(\ref{eq:nqb-2}), we obtain
\begin{eqnarray}
\lefteqn{
n(-l)\cdot Q_{\mathrm{B}} \Phi} \nonumber\\
&&
   = e^{-\frac{\pi^{2}}{2\epsilon} |l|}
   l \frac{\partial}{\partial l} \left(
     e^{\frac{\pi^{2}}{2\epsilon} |l|}
     |l| \frac{\mathcal{N}^{2}}{4\epsilon} \bchi (l) \right)
= \left\{
    \begin{array}{ll}
      \displaystyle
      \frac{\mathcal{N}^{2}}{4\epsilon}
      l \left( \frac{\partial}{\partial l}
                 + \frac{\pi^{2}}{2\epsilon}
      \right) \left( l \chi (l) \right)
          & \mbox{for $l>0$} \\[1.5ex]
      \displaystyle
      -\frac{\mathcal{N}^{2}}{4\epsilon} l \left(
        \frac{\partial}{\partial l} - \frac{\pi^{2}}{2\epsilon} \right)
        \left( l \bar{\chi}(-l) \right)
           & \mbox{for $l<0$}
     \end{array}
    \right.~.
    \label{eq:nqb-3}
\end{eqnarray}

Second, we consider the non-linear term
$g\left| \Phi \ast \Phi \right\rangle$.
Keeping the most dominant terms in the limit 
$\epsilon\rightarrow 0$, we find that
\begin{eqnarray}
\lefteqn{
n(-l)\cdot \left(\Phi \ast \Phi \right)
= \int d1 d2 d3 d4\,
    \left\langle V_{3} (1,2,3) \left. \right|
    \Phi \right\rangle_{1}\left| \Phi \right\rangle_{2}
    \, {}_{4} \! \left\langle n(-l) 
    \left| R(3,4) \right\rangle\right.
  } \nonumber\\
&&= -\int d1d2d3\, \left\langle V_{3}(1,2,3)\right|
     \left. \Phi \right\rangle_{1}\left| \Phi \right\rangle_{2}
     \left| n(-l) \right\rangle_{3}
    \nonumber\\
&&= -\int d1d2d3\, \left\langle V_{3}^{0} (1,2,3) \right|
     C^{(r)}(\rho_{I})
     \left| \Phi \right\rangle_{1}\left| \Phi \right\rangle_{2}
     \left| n(-l) \right\rangle_{3}  \nonumber\\
&&= - \int^{\infty}_{-\infty} dl_{1}dl_{2} \int d1d2d3\,
   \left\langle V_{3}^{0} (1,2,3) \right|
     \Big[ C^{(r)}(\rho_{I}) \bar{\pi}^{(1)}_{0}
       \left| n(l_{1}) \right\rangle_{1}
       \left| n(l_{2}) \right\rangle_{2}
       \left| n(-l) \right\rangle_{3} \bchi (l_{1}) \bphi (l_{2})
    \nonumber\\
&& \hspace{12.5em}
        {}+C^{(r)}(\rho_{I})\bar{\pi}^{(2)}_{0}
       \left| n(l_{1}) \right\rangle_{1}
       \left| n(l_{2}) \right\rangle_{2}
       \left| n(-l) \right\rangle_{3} 
       \bphi (l_{1}) \bchi (l_{2}) \Big]\nonumber\\
&& \qquad \quad {}+ \cdots~,
     \label{eq:nphiphi-1}
\end{eqnarray}
where $r$ of $C^{(r)}$ can be any of $1,2,3$. 
In going from the third line to the fourth line in the 
above equation, we expand $|\Phi\rangle_{1,2}$ in terms 
of the complete basis defined in subsection \ref{sec:expansion}. 
In doing so, we have used the following idempotency equations
\begin{equation}
\int d3 \langle V_3^0(1,2,3)|n(-l)\rangle_3
\propto
\int dl_1 \int dl_2 \delta (l_1+l_2-l)
\!\;_1\langle n(l_1)| \!\;_2 \langle n(l_2)|\frac{1}{\alpha_1\alpha_2}~,
\end{equation}
for $|\alpha_1|,|\alpha_2|<|l|$, and 
\begin{equation}
\int drds\langle V_3^0(1,2,3)|n(l_r)\rangle_r |n(l_s)\rangle_s
\propto
\!\;_t\langle n(-l_r-l_s)|\frac{1}{\alpha_t}~,
\label{rst}
\end{equation}
where $r,s,t\in \{1,2,3\}$ and $r>s,~r\neq t,~s\neq t$. 
These equations hold in the limit $\epsilon\rightarrow 0$ and 
can be proved by using the connection conditions satisfied by 
$\langle V_3^0|$. 
`$\cdots$' on the rightest hand side
stands for the contributions from the component fields
other than $\bphi (l)$ and $\bchi (l)$. 
One can easily see that these terms include one 
annihilation operator other than $\phi ,\chi$ and 
one creation operator other than $\bar{\phi}, \bar{\chi}$. 
We will ignore these contributions in the rest of this
appendix.
Combining the ghost number conservation
with the above arguments,
we obtain the last equality in eq.(\ref{eq:nphiphi-1}).
Since we may choose an arbitrary $r$ for $C^{(r)}$ on
the three string vertex $\left\langle V_{3}^{0} \right|$
as mentioned above,
we set $r=1$ in the first term and set $r=2$ in the second term
on the rightest hand side in eq.(\ref{eq:nphiphi-1}).
{}From the definition (\ref{eq:n0-1}) of the state$|n(l)\rangle$
and  the fact that
the field $C(\rho_{I})$ satisfy the Dirichlet boundary condition
on the state $\left|B_{0} \right\rangle$,
one can find that
\begin{equation}
C\left(\rho_{I}\right)\left|n(l)\right\rangle
 = \mathcal{O}(\epsilon)~,
\end{equation}
and thus
\begin{equation}
C\left(\rho_{I}\right) \bar{\pi}_{0} \left| n(l) \right\rangle
 = \left| n(l) \right\rangle + \mathcal{O}(\epsilon)~.
\end{equation}
This implies that
in the leading order of $\epsilon$,
eq.(\ref{eq:nphiphi-1}) becomes
\begin{eqnarray}
&&\label{eq:eq:nphiphi-2}
n(-l)\cdot \left(\Phi \ast \Phi \right) \\
&& = -  \int^{\infty}_{-\infty} dl_{1} dl_{2}
    \int d1d2d3\,  \left\langle V_{3}^{0} (1,2,3) \right|
      \left. n(l_{1})\right\rangle_{1}
      \left| n(l_{2}) \right\rangle_{2}
      \left| n(-l) \right\rangle_{3}
     \left( \bchi(l_{1}) \bphi (l_{2}) 
           + \bphi(l_{1})\bchi(l_{2})\right)
   \nonumber\\
&& = -\pi^{3}
   \frac{\left(4\pi^{2}\palpha\right)^{\frac{p+1}{2}}}
        {\left( 2\pi^{2} \palpha\right)^{\frac{13}{2}}}
   \sqrt{\frac{V_{\mathrm{D}}}{V_{\mathrm{N}}}}
   \frac{\mathcal{N}^{3}}{\epsilon^{3}}
    \int^{\infty}_{-\infty} dl_{1} dl_{2}\,
        \left|l_{1}l_{2}l\right|
        \delta\left( l_{1}+l_{2}-l \right) \frac{1}{2}
        \left( \bchi (l_{1})\bphi(l_{2})+\bphi(l_{1})\bchi(l_{2})
          \right)~.\nonumber
\end{eqnarray}
In the second equality in this equation, we have used
the result obtained in Appendix~\ref{sec:liouville}.
We can further recast eq.(\ref{eq:eq:nphiphi-2}) into
\begin{eqnarray}
\lefteqn{
 n(-l) \cdot \left( \Phi \ast \Phi \right)} 
\nonumber\\
  &&= - \pi^{3}
   \frac{\left(4\pi^{2}\palpha\right)^{\frac{p+1}{2}}}
        {\left( 2\pi^{2} \palpha\right)^{\frac{13}{2}}}
   \sqrt{\frac{V_{\mathrm{D}}}{V_{\mathrm{N}}}}
   \frac{\mathcal{N}^{3}}{\epsilon^{3}}\;  |l|
  \int^{\infty}_{-\infty} dl_{1}\,
    \left| l_{1}(l-l_{1}) \right|
    \bchi(l_{1}) \bphi (l-l_{1}) \nonumber\\
&&= -  \pi^{3}
   \frac{\left(4\pi^{2}\palpha\right)^{\frac{p+1}{2}}}
        {\left( 2\pi^{2} \palpha\right)^{\frac{13}{2}}}
   \sqrt{\frac{V_{\mathrm{D}}}{V_{\mathrm{N}}}}
   \frac{\mathcal{N}^{3}}{\epsilon^{3}}\; |l| \times
   \nonumber\\
 && \quad\quad \times
    \left[ \Theta(l) \left\{
      \int^{l}_{0} dl_{1}\,
          l_{1}(l-l_{1}) \chi (l_{1}) \phi(l-l_{1})
      \right.\right. \nonumber\\
      &&\hspace{6em}
      \left.
    {}+\int^{\infty}_{0} dl_{1}\,
              l_{1}(l_{1}+l)\Big(
                 \chi(l+l_{1}) \bar{\phi}(l_{1})
          + \bar{\chi}(l_{1}) \phi(l+l_{1}) \Big)
       \right\}
    \nonumber\\
&& \qquad\quad {} + \Theta (-l)
    \left\{ \int^{-l}_{0} dl_{1}\,  l_{1}(-l-l_{1})
             \bar{\chi}(l_{1}) \bar{\phi}(-l-l_{1})
    \right. \nonumber\\
   && \hspace{7em}\left. \left.
    +\int^{\infty}_{0}dl_{1}\,
       l_{1}(l_{1}-l)
           \Big( \bar{\chi}(-l+l_{1})\phi(l_{1}) 
           + \chi(l_{1})\bar{\phi}(l_{1}-l)
             \Big)  \right\}
    \right]~,\label{eq:rhs-brs2}
\end{eqnarray}
where $\Theta$ is the step function
defined as $\Theta (x)=1$ for $x>0$ and $=0$ for $x<0$.

Combining eqs.(\ref{eq:brs-lhs}), (\ref{eq:nqb-3})
and (\ref{eq:rhs-brs2}), we obtain eq.(\ref{eq:brst}).

\section{$\int d1d2d3\left\langle V^0_3(1,2,3)\right|
 \left.n(l_1)\right\rangle_1
\left|n(l_2)\right\rangle_2  \left|n(-l)\right\rangle_3$}
\label{sec:liouville}

\begin{figure}[htbp]
\begin{center}
\includegraphics[width=30em,clip]{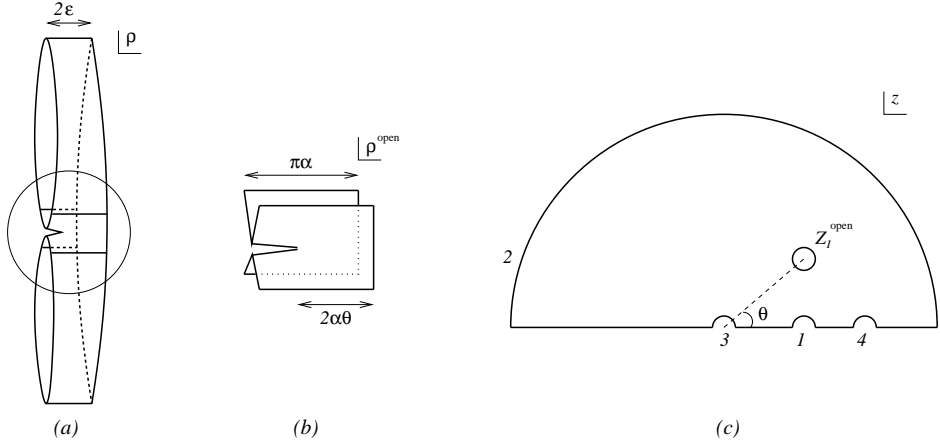}
\begin{quote}
\caption{(a) The closed string 3-point diagram 
in the limit $\epsilon \to 0$.
(b) The open string 4-point diagram.
(c) The upper half $z$-plane after 
cutting out the circle 
around the interaction point $Z_I$ 
and semicircles around the points $Z_r$.}
\label{fig:open}
\end{quote}
\end{center}
\end{figure}

In this appendix,
we will prove that the following relation holds
in the leading order of $\epsilon$:
\begin{eqnarray}
\lefteqn{\int d1d2d3\left\langle V^0_3(1,2,3)\right|
 \left.n(l_1)\right\rangle_1
\left|n(l_2)\right\rangle_2  \left|n(-l)\right\rangle_3}
\nonumber\\
&&\simeq\frac{1}{2}
\left(\pi^3\frac{(4\pi^2\alpha^\prime)^{\frac{p+1}{2}}}
{(2\pi^2\alpha^\prime)^{\frac{13}{2}}}
\sqrt{\frac{V_D}{V_N}}\right)
\frac{\mathcal{N}^3}{\epsilon^3}|l_1l_2l|\delta(l_1+l_2-l)~.
\label{eq:vnnn}
\end{eqnarray}
This equation was used in eq.(\ref{eq:eq:nphiphi-2})
to derive eq.(\ref{eq:brst}).
Therefore, by proving eq.(\ref{eq:vnnn}),
we can complete the derivation of eq.(\ref{eq:brst})
presented in the last appendix.
{}From the definition (\ref{eq:n0-1}),
we find that we can prove the above equation
by evaluating
\begin{eqnarray}
\int d'1d'2d'3\left\langle V^{0}_3(1,2,3)\right| 
\left. B_0\right\rangle_1^\epsilon 
\left|B_0\right\rangle_2^\epsilon
 \left|B_0\right\rangle_3^\epsilon~.
\label{eq:vbbb}
\end{eqnarray}
Here we have introduced the integration measure $d'r$
for zero modes defined by removing the $\alpha$ dependence from
$dr$ given in eq.(\ref{eq:zeromodemeasure}):
\begin{equation}
d'r=(2\pi)^{-27}d^{26}p_rid\bar{\pi}_0^{(r)}d\pi_0^{(r)}~.
\end{equation}
The amplitude (\ref{eq:vbbb}) corresponds to the
string diagram described in Fig.~\ref{fig:open}(a).

As was performed in 
Ref.\cite{Kishimoto:2004jk},
we use conformal field theory (CFT) technique to calculate 
eq.(\ref{eq:vbbb}). 
In the $OSp$ invariant string theory, 
the ghost sector $(C,\bar{C})$ is described by the CFT with 
central charge $-2$. The full theory, which
consists of the matter sector $X^{\mu}$
in addition to the ghost sector,
is therefore the CFT with total central charge $c=24$. 
Since $c\neq 0$, the amplitudes of this system depend
on the metric on the worldsheet. 
The CFT we are dealing with consists of 
free bosons and fermions. 
Therefore the metric dependence stems from 
the determinant of the Laplacian on the worldsheet. 
It can be given by evaluating the Liouville 
action on the worldsheet \cite{Alvarez:1982zi}\cite{GSW}. 

\begin{figure}[htbp]
\begin{center}
\includegraphics[width=26em,clip]{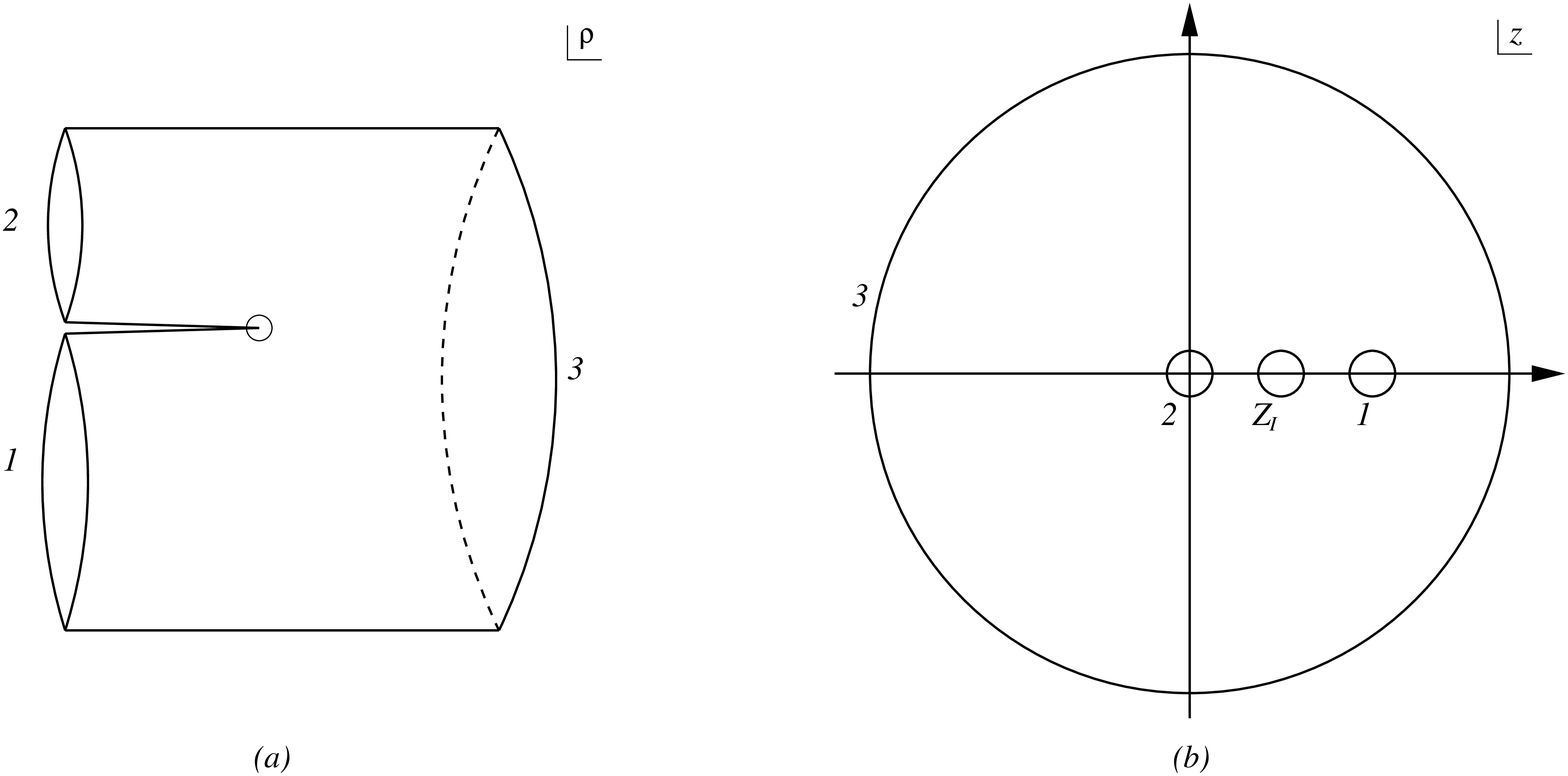}
\begin{quote}
\caption{(a) The closed string 3-point diagram.
(b) The complex $z$-plane after 
cutting out circles 
around the interaction point $Z_I$ 
and the points $Z_r$. }
\label{fig:czplane}
\end{quote}
\end{center}
\end{figure}

The oscillator independent part of the three string vertex
$\langle V_{3}^0|$ can be considered to be due to the contributions 
from the Liouville action. 
$\langle V_{3}^0|$ corresponds to the diagram 
depicted in Fig.~\ref{fig:czplane}(a).
It is useful to pull this diagram ($\rho$-plane)
back to the complex $z$-plane
(Fig.~\ref{fig:czplane}(b))
through the Mandelstam mapping
\cite{Mandelstam:1973jk},
\begin{equation}
\rho(z)=\alpha_1\ln(z-Z_1)
+\alpha_2\ln(z-Z_2)+\alpha_3\ln(z-Z_3)~.
\label{eq:mandelstam}
\end{equation}
We fix the $SL(2,\mathbb{C})$ gauge symmetry
on the worldsheet by setting 
$Z_1=1$, $Z_2=0$ and $Z_3=\infty$.
The interaction point is at $Z_I=-\frac{\alpha_2}{\alpha_3}$,
where we have $\frac{\partial\rho}{\partial z}=0$.
In order to avoid the singularity at the interaction point,
we cut out a small circle of radius $r_I$ around 
the interaction point in the $\rho$-plane.
We also cut the points corresponding to 
incoming and outgoing strings at $\tau=\pm\infty$
by terminating each string at $\tau=\tau_r$
$(r=1,2,3)$.
These correspond to cutting circles out of 
the $z$-plane centered on $z=Z_{p}$ 
with small radii $\varepsilon_p$ $(p=1,2,3,I)$,
as represented in Fig.~\ref{fig:czplane}(b).

Let us suppose that
the $\rho$-plane is equipped with the flat metric:
\begin{equation}
ds^{2}=d\rho d\bar{\rho}
=e^{\phi}dzd\bar{z}~,
\quad
\phi=\ln\left|\frac{\partial\rho}{\partial z}\right|^2~.
\label{rho-metric}
\end{equation}
We will see that $\langle V_{3}^0|$ is constructed to reproduce 
the CFT amplitudes on the $\rho$-plane with this metric. 
Since the calculations are done on the $z$-plane, 
we should take care of the Liouville action 
for the Liouville field $\phi$ in eq.(\ref{rho-metric}), 
because the determinant of the Laplacian is expressed as 
\begin{equation}
\left.\ln {\det}^\prime \Delta\right|_{\phi}
-\left.\ln {\det}^\prime \Delta\right|_{\phi=0}
\sim -\frac{1}{48\pi}\left[\int 
         d^2\sigma\partial_a\phi\partial^a\phi
+4\int_{\partial\mathcal{M}} ds\, \hat{k}\phi\right]~,
\end{equation} 
where $s$ denotes the variable parametrizing
the boundary of the worldsheet $\mathcal{M}$;
$\hat{k}$ denotes the geodesic curvature of the boundary
defined as
\begin{equation}
\hat{k}=n_b t^a \hat{\nabla}_a t^b~,
\end{equation}
where $t^{a}$ is the unit vector tangential to
the boundary while
$n^{a}=-\frac{\epsilon^{ab}}{\sqrt{\hat{h}}}t_{b}$
is normal,
and $\hat{\nabla}_a$ denotes the covariant derivative
associated with the metric $ds^2=dzd\bar{z}$.

The dependence of 
$\left.\ln {\det}^\prime \Delta\right|_{\phi=0}$
on  $\varepsilon_p$
was calculated in Appendix 11.A of Ref.\cite{GSW}
\footnote{Eq.(\ref{eq:epsilon-dependence})
          is twice eq.(11.A.26) in Ref.\cite{GSW}
          because we are dealing with the closed string case. 
}:
\begin{equation}
\left.\ln {\det}^\prime \Delta\right|_{\phi=0}
\sim-\frac{1}{3}\sum_{p}\ln\varepsilon_p~,
\label{eq:epsilon-dependence}
\end{equation}
where $\sum_p$ denotes the sum over all the values of $I$ and $r$.
By exploring the Mandelstam mapping (\ref{eq:mandelstam})
near the cuts,
we find that $\varepsilon_p$ depend on $\alpha_r$
as follows:
\begin{equation}
\ln\varepsilon_{r}
 \sim {}-\tau_{r}+\frac{\hat{\tau}_0}{\alpha_{r}}
\quad (r=1,2,3)~, \qquad
\ln\varepsilon_I \sim \frac{1}{2}(\ln 2r_I-\ln|c_I|)~,
\end{equation}
where 
\begin{equation}
c_I=\left.\frac{\partial^2\rho}{\partial z^2}\right|_{z=Z_I}
=\frac{{\alpha_3}^3}{\alpha_1\alpha_2}~.
\end{equation}
Using these results, we obtain
\begin{eqnarray}
\ln {\det}^\prime \Delta
&\sim&
\frac{1}{6}\sum_{r=1}^3\frac{\hat{\tau}^0}{\alpha_r}
+\frac{1}{12}\sum_{r=1}^3\ln|\alpha_r|~.
\end{eqnarray}
Thus we find that
the determinant factor 
depends on $\alpha_r$ in the following way,
\begin{eqnarray}
\left({\det}^\prime \Delta\right)^{-\frac{c}{2}}
=\left({\det}^{\prime} \Delta \right)^{-12}
\propto
|\mu(1,2,3)|^2\frac{1}{|\alpha_1\alpha_2\alpha_3|}~.
\label{oscillatorindep}
\end{eqnarray}
This reproduces the factor appearing in the three
string vertex $\langle V_{3}^0|$ given in eq.(\ref{V30}),
except for the fact that we should take the absolute value 
of the factor $\alpha_1\alpha_2\alpha_3$. 
Thus we see that roughly speaking,
 $\langle V_{3}^0|$ is constructed to reproduce 
the CFT amplitudes on the $\rho$-plane with the metric
(\ref{rho-metric}). 
Precisely speaking, 
$\mathrm{sgn} \left(\alpha_{1}\alpha_{2}\alpha_{3}\right)
  \langle V^{0}_{3} |$ corresponds to Fig.~\ref{fig:czplane}(a).

Now we would like to
calculate eq.(\ref{eq:vbbb}) in the limit $\epsilon\to 0$.
For this purpose, it is convenient  to
see the string diagram Fig.~\ref{fig:open}(a)
{}from the point of view of 
the dual open string channel.
In this channel, one can regard
the worldsheet as being
swept by four open strings interacting 
via mid-point type interaction
Fig.~\ref{fig:open}(b).
In the limit $\epsilon\to 0$,
such open strings propagate through
very long proper time and thus
the most dominant contribution comes from
the propagations of the open string tachyons.
The propagator contributes the factor
\begin{equation}
e^{\frac{\pi^2}{\epsilon}
          \mathrm{max}(|\alpha_1|,|\alpha_2|,|\alpha_3|)}~.
\label{eq:leg}
\end{equation}

Let us evaluate 
the determinant of the Laplacian
on the worldsheet depicted by Fig.~\ref{fig:open}(b).
The Mandelstam mapping from the upper half $z$-plane 
into the open string 4-point 
$\rho^{\mathrm{open}}$-plane is
\begin{eqnarray}
\lefteqn{
\rho^{\mathrm{open}}(z)=
\alpha_1^{\mathrm{open}}\ln(z-Z_1^{\mathrm{open}})
+\alpha_2^{\mathrm{open}}\ln(z-Z_2^{\mathrm{open}})}\nonumber\\
&&\hspace{4.5em}
+\alpha_3^{\mathrm{open}}\ln(z-Z_3^{\mathrm{open}})
+\alpha_4^{\mathrm{open}}\ln(z-Z_4^{\mathrm{open}})~.
\end{eqnarray}
For the worldsheet described in Fig.~\ref{fig:open}(b),
we should choose
\begin{equation}
\alpha_1^{\mathrm{open}}=\alpha_2^{\mathrm{open}}
=-\alpha_3^{\mathrm{open}}=-\alpha_4^{\mathrm{open}}
=\frac{2\epsilon}{\pi}\equiv\alpha~.
\end{equation}
Let us choose $Z^{\mathrm{open}}_r
                 =(1,\infty,0,x)$ $(r=1,2,3,4)$.
Here we should take $x>1$ to treat the worldsheet
that we are considering.
The interaction point is 
\begin{equation}
Z_I^{\mathrm{open}}=1+i\sqrt{x-1}~,
\end{equation}
which is a solution of 
$\left.\frac{d\rho^{\mathrm{open}}}{dz}
     \right|_{z=Z_I^{\mathrm{open}}}=0$.
We introduce the parameter $\theta$ defined by
\begin{equation}
\cos\theta = \frac{1}{\sqrt{x}}~,\quad
\sin\theta  = \frac{\sqrt{x-1}}{\sqrt{x}}~.
\end{equation}
In terms of $\theta$, the interaction point is described by
\begin{equation}
Z_I^{\mathrm{open}}=\frac{1}{\cos\theta}e^{i\theta}~,
\quad
\rho^{\mathrm{open}}_I\equiv
\rho^{\mathrm{open}}(Z_I^{\mathrm{open}})
=2\alpha(\ln\cos\theta-i\theta)~.
\end{equation}
$\theta=\frac{\pi}{4}$ in our case, but
let us treat $\theta$ as a free parameter 
in order to compare with the results of \cite{Kishimoto:2003ru}.
In order to avoid the singularities,
we excise small circles around the interaction point
and external strings, 
as shown in Fig.~\ref{fig:open}(c). 
We define $\tau_r^{\mathrm{open}}$ accordingly. 
%
%
%
%
By using the metric (\ref{rho-metric}),
we find that the
moduli dependence of the determinant of the Laplacian
becomes
\begin{eqnarray}
\ln{\det}^\prime\Delta
&\sim&
-\frac{1}{48\pi}\left[\int d^2\sigma\partial_a\phi\partial^a\phi
+4\int_{\partial\mathcal{M}}ds\, k\phi\right]
+\frac{1}{6}\sum_{r=1}^4\ln\varepsilon_r^{\mathrm{open}}
+\frac{1}{3}\ln\varepsilon_I^{\mathrm{open}}\nonumber\\
&\sim& 
\frac{1}{4}\ln\left(\alpha \cos\theta \sin\theta\right)~.
\label{eq:lndetprime}
\end{eqnarray}
Here we have used 
\begin{eqnarray}
\ln\varepsilon_1^{\mathrm{open}} &\sim& 
  -\tau^{\mathrm{open}}_1+\frac{\hat{\tau}_0^{\mathrm{open}}}{\alpha}
   +\ln(x-1)~,\nonumber\\
\ln\varepsilon_2^{\mathrm{open}} &\sim& 
   -\tau^{\mathrm{open}}_2+\frac{\hat{\tau}_0^{\mathrm{open}}}{\alpha}
   -\ln x~,\nonumber\\
\ln\varepsilon_3^{\mathrm{open}} &\sim& 
   -\tau^{\mathrm{open}}_3
   -\frac{\hat{\tau}_0^{\mathrm{open}}}{\alpha}~,\nonumber\\
\ln\varepsilon_4^{\mathrm{open}} &\sim& 
   -\tau^{\mathrm{open}}_4
   -\frac{\hat{\tau}_0^{\mathrm{open}}}{\alpha}-\ln x+\ln(x-1)~,
   \nonumber\\
\ln\varepsilon_I^{\mathrm{open}}  &\sim& 
   \frac{1}{2}(\ln 2r_I^{\mathrm{open}} - \ln|c_I^{\mathrm{open}}|)~,
\end{eqnarray}
with
\begin{eqnarray}
c_I^{\mathrm{open}}&=&\frac{2i\alpha}{\sqrt{x-1}(1+i\sqrt{x-1})^2}
=\frac{2i\alpha\cos^3\theta}{e^{2i\theta}\sin\theta}~,
\nonumber\\
\hat{\tau}_0^{\mathrm{open}}
&=&
2\alpha \ln \cos \theta ~.
\end{eqnarray}
Combining eqs.(\ref{eq:leg}) and (\ref{eq:lndetprime}),
we find that the amplitude corresponding to
the pants diagram
in the limit $\epsilon\rightarrow 0$ depends on 
$\alpha_r$, $\alpha$ and $\theta$ as 
\begin{eqnarray}
\int d'1d'2d'3 \left\langle V^0_3(1,2,3)\right| 
\left.B_0\right\rangle_1^{(\pi-2\theta)\alpha}
\left|B_0\right\rangle_2^{(\pi-2\theta)\alpha}
\left|B_0\right\rangle_3^{2\theta\alpha}
&\propto&
\frac{e^{\frac{\pi^2}{\epsilon}
\mathrm{max}(|\alpha_1|,|\alpha_2|,|\alpha_3|)}}
{\alpha^3\sin^3\theta\cos^3\theta}~.
\end{eqnarray}

In Ref.\cite{Kishimoto:2003ru}, 
the authors computed this quantity
in the limit $\theta\to 0$,
by using the Cremmer-Gervais identity \cite{Cremmer:1974ej}.
By comparing our result with theirs,
it is straightforward to determine the overall constant,
and we obtain
\begin{eqnarray}
\lefteqn{\int d'1d'2d'3\left\langle V^0_3(1,2,3)\right|
\left.B_0\right\rangle_1^\epsilon
\left|B_0\right\rangle_2^\epsilon
\left|B_0\right\rangle_3^\epsilon}
\nonumber\\
&&\simeq\frac{1}{2}\left(\pi^3
\frac{(4\pi^2\palpha)^{\frac{p+1}{2}}}
{(2\pi^2\palpha)^{\frac{13}{2}}}
\sqrt{\frac{V_D}{V_N}}\right)
\frac{\mathcal{N}^3}{\epsilon^3}
\frac{|\alpha_1\alpha_2\alpha_3|}{\alpha_1\alpha_2\alpha_3}
e^{\frac{\pi^2}{\epsilon}
          \mathrm{max}(|\alpha_1|,|\alpha_2|,|\alpha_3|)}~.
    \label{eq:vbbb-result}
\end{eqnarray}
When we substitute eq.(\ref{eq:vbbb-result})
into eq.(\ref{eq:vnnn}),
the exponential factor on the right hand side
in this equation is canceled 
by the regularization factors
$e^{-\frac{\pi^{2}}{2\epsilon} |l_{r}|}$
$(r=1,2,3)$ in $|n(l_{r})\rangle_{r}$.
It is because
$\mathrm{max}(|\alpha_1|,|\alpha_2|,|\alpha_3|)
 =\frac{1}{2}(|\alpha_1|+|\alpha_2|+|\alpha_3|)$.
Finally, carrying out the $\alpha_r$ integration,
we obtain eq.(\ref{eq:vnnn}).

A comment is in order. 
In eq.(\ref{eq:vnnn}), there is a factor 
$|l_1l_2l|$. The absolute value originates from the one in 
eq.(\ref{oscillatorindep}).  
Taking the absolute value is necessary 
to be consistent with 
\begin{eqnarray}
& &
\left(\int d1d2d3 
\langle V_3^0(1,2,3)|n(l_1)\rangle_1|n(l_2)\rangle_2|n(l_3)\rangle_3
\right)^{\dagger}
\nonumber
\\
& &
=
\int d1d2d3 
\langle V_3^0(1,2,3)|n(-l_1)\rangle_1|n(-l_2)\rangle_2|n(-l_3)\rangle_3 ~,
\end{eqnarray}
which can be easily proved. 


\end{document}